\documentclass[ 
    aps,                
    prd,                
    twocolumn,           
    tightenlines,       
    12pts,              
    groupedaddress,     
    superscriptaddress, 
    preprintnumbers,    
    floatfix,
    nofootinbib,        
    titlepage,          
    showkeys,           
    showpacs,           
    ]{revtex4-2}
\usepackage[utf8]{inputenc}
\usepackage{epsfig,amsfonts,amsthm}
\usepackage{epstopdf}
\usepackage{amsfonts,amsmath,amssymb}
\usepackage{latexsym}
\usepackage{bm}             
\usepackage{bbm}
\usepackage{array}
\usepackage{comment}
\usepackage{hyperref}
\usepackage[compat=1.1.0]{tikz-feynman} 	
\usepackage{color}
\usepackage{pb-diagram}
\usepackage{slashed}
\usepackage{multirow}
\usepackage{cancel}
\usepackage{placeins}                       
\usepackage{diagbox}

\newcommand{\be}{\begin{equation}}
\newcommand{\ee}{\end{equation}}
\newcommand{\bea}{\begin{eqnarray}}
\newcommand{\eea}{\end{eqnarray}}

\def\eff{ \textrm{eff} }

\def\apriori{{\em a priori }}
\def\lra{\leftrightarrow}
\def\ra{\rightarrow}

\def\lsim{\mathrel{\rlap{\lower4pt\hbox{\hskip1pt$\sim$}}
    \raise1pt\hbox{$<$}}}                
\def\gsim{\mathrel{\rlap{\lower4pt\hbox{\hskip1pt$\sim$}}
    \raise1pt\hbox{$>$}}}                

\def\S{{\scriptscriptstyle \mathrm{S} }}
\def\H{{\scriptscriptstyle \mathrm{H} }}
\def\HS{{\scriptscriptstyle \mathrm{HS} }}
\def\N{{\scriptscriptstyle \mathrm{N} }}

\def\N{{\scriptscriptstyle \mathrm{N} }}

\def\I{{\scriptscriptstyle \mathrm{I} }}

\newcommand{\Mol}{\textrm{M\o l}}

\def\N{{\scriptscriptstyle \rm N}}

\usepackage{color}
\definecolor{myorange}{rgb}{1,0.5,0}

\begin{document}
%

\title{Momentum distributions of cosmic relics: Improved analysis}

\author{Kalle Ala-Mattinen}
\email{kalle.ala-mattinen@helsinki.fi}
\affiliation{Department of Physics, University of Helsinki, 
                      P.O.Box 64, FI-00014 University of Helsinki, Finland}
\affiliation{Helsinki Institute of Physics, 
                      P.O.Box 64, FI-00014 University of Helsinki, Finland}

\author{Matti Heikinheimo}
\email{matti.heikinheimo@helsinki.fi}
\affiliation{Department of Physics, University of Helsinki, 
                      P.O.Box 64, FI-00014 University of Helsinki, Finland}
\affiliation{Helsinki Institute of Physics, 
                      P.O.Box 64, FI-00014 University of Helsinki, Finland}

\author{Kimmo Kainulainen}
\email{kimmo.kainulainen@jyu.fi}
\affiliation{Department of Physics, University of Jyv\"askyl\"a}
\affiliation{Helsinki Institute of Physics, 
                      P.O.Box 64, FI-00014 University of Helsinki, Finland}

\author{Kimmo Tuominen}
\email{kimmo.i.tuominen@helsinki.fi}
\affiliation{Department of Physics, University of Helsinki, 
                      P.O.Box 64, FI-00014 University of Helsinki, Finland}
\affiliation{Helsinki Institute of Physics, 
                      P.O.Box 64, FI-00014 University of Helsinki, Finland}

\begin{abstract}

\noindent
We solve coupled momentum-dependent Boltzmann equations for the phase space distribution of cosmic relic particles, without resorting to approximations of assuming kinetic equilibrium or neglecting backscattering or elastic interactions. Our method is amendable to precision numerical computations. To test it, we consider two benchmark models where the momentum dependence of dark matter distribution function is potentially important: a real singlet scalar extension near the Higgs resonance and a sterile neutrino dark matter model with a singlet scalar mediator. The singlet scalar example shows that the kinetic equilibrium may hold surprisingly well even near sharp resonances.
However, the integrated method may underestimate the relic density by up to 40\% in extreme cases. In the sterile neutrino dark matter model, we studied how the inclusion of previously ignored elastic interactions and processes with initial state sterile neutrinos 
could affect the nonthermal nature of their resulting distributions. 
Here the effects turned out to be negligible, proving the robustness of the earlier predictions.
\end{abstract}

\preprint{HIP-2021-37/TH}
\maketitle

%
\section{Introduction}  
%

Current cosmological observations can be accommodated within the cold dark matter (CDM) paradigm~\cite{Planck:2015fie}. This hypothesis is appealing 
within our present understanding of the structure of ordinary matter: extending the Standard Model (SM) of elementary particles and their interactions  with dark matter particle degrees of freedom allows the abundance of CDM to be created by thermal production and decoupling of the dark matter particles  in the expanding early Universe. The standard treatment for calculating this abundance of CDM particles relies on the Zel'dovich-Okun-Pikelner-Lee-Weinberg (ZOPLW) equation~\cite{Zeldovich:1965,Lee:1977ua}. 

The ZOPLW equation is obtained from the Boltzmann equation by integrating over the phase space of the dark matter under the assumptions of detailed balance and kinetic equilibrium distributions~\cite{Gondolo:1990dk, Griest:1990kh,Binder:2017rgn}. However, the momentum distribution of dark matter may contain essential information that is neglected in this treatment. For example, if the dark matter production takes place at a resonance region, where the DM annihilation rate is strongly momentum dependent, the elastic reactions might not be fast enough to keep kinetic equilibrium. In this setting the true annihilation rate and hence the final DM abundance may deviate from the value obtained under the equilibrium assumption~\cite{Ala-Mattinen:2019mpa,Binder:2017rgn,Abe:2021jcz}. Another example concerns warm dark matter (WDM) whose momentum distribution may directly influence the cosmic structure formation by reducing the number of DM halos at small scales compared to CDM. In kinetic equilibrium, the suppression of the matter power spectrum can be well approximated via a single scale given by the WDM mass~\cite{Drewes:2016upu}. However, if the DM particle is not in kinetic equilibrium, the resulting suppression may be more complicated~\cite{Merle:2015oja,Konig:2016dzg,Dienes:2020bmn,Drewes:2016upu}.

Earlier calculations accounting for the DM momentum distributions tend to rely on simplifying approximations. For example, in the analysis of~\cite{Merle:2015oja,Konig:2016dzg}, the elastic interactions of the initial state DM particles have been neglected. 
In this paper, we complement these earlier analyses by presenting a numerical method based on discretization of the momentum space that allows for a completely general solution of the Boltzmann equations for the momentum distributions of multiple number of particle species, any number of which can be out of equilibrium. We note that advanced momentum-dependent methods have also been developed and used to treat 
neutrino oscillations in the early Universe~\cite{Kainulainen:2001cb,Ghiglieri:2015jua,Hannestad:2015tea,Bodeker:2020hbo}.

We demonstrate our method in the context of the dark matter production near a sharp resonance, comparing our results with the ZOPLW approach and with the momentum-dependent method of~\cite{Ala-Mattinen:2019mpa}, which uses a generalized relaxation time approximation for the numerically expensive backreaction terms. Our results validate the approximation scheme of~\cite{Ala-Mattinen:2019mpa} to its expected accuracy. 
Furthermore, we find that this scheme slightly overestimates the effect of elastic scattering channels, and this seems to be the case also with the truncated derivative methods used in \cite{Binder:2017rgn,Abe:2021jcz}.
We also apply our method in a sterile neutrino DM model including a singlet scalar mediator, first analyzed with simplified evolution equations in~\cite{Merle:2015oja,Konig:2016dzg}. We find that neither including elastic interaction channels, nor adding new collision terms induced by a symmetry breaking changes the results appreciably. This verifies that the approximations used in~\cite{Merle:2015oja,Konig:2016dzg} 
are robust and their results remain valid in the full solution. 

The paper is organized as follows: In Sec.~\ref{sec:kinetic_equation} we describe the discretization of the collision integrals. We  then apply the developed methodology to two benchmark models. first, in Sec.~\ref{sec:modelSS} to the model where SM is extended with a real singlet scalar and then in Sec.~\ref{sec:modelSSN} to the model where the additional fields are a real singlet scalar and a sterile neutrino. In Sec.~\ref{sec:checkout} we present our conclusions and outlook toward further work. Many details of the computations can be found in the Appendixes.

%
\section{The kinetic equation}  
\label{sec:kinetic_equation}
%

In an expanding homogeneous and isotropic universe the Boltzmann equation can be written as
\begin{equation}
\label{eq:BE_line1}
	\frac{\partial f(p,t)}{\partial t} - Hp\frac{\partial f}{\partial p}
	=
	\sum \mathcal{C}_{\textrm{coll}}[f] \,,
\end{equation}
where $\mathcal{C}_{\textrm{coll}}$ are the collision terms describing the chemical and kinetic balances, $H = \dot{a}/a$ is the Hubble parameter and $a(t)$ is the scale factor. Expansion of the Universe is best quantified by integrating along the curves of constant comoving momentum, $k = ap$. Then, writing the Liouville operator in terms of $k$, the momentum derivative vanishes and we have
\begin{equation}
\frac{d f(k,t)}{d t}
	=
	\sum \mathcal{C}_{\textrm{coll}}[f] \,,
\end{equation}
where we identified $f(p,t) = f(k/a(t),t) \equiv f(k,t)$. It is more natural to work with temperature instead of time, thus we define a dimensionless variable $x \equiv m_0/T$, where $m_0$ is some reference scale, and $T$ is photon temperature. To evaluate the Jacobian of this transformation, we use the adiabatic radiation era time-temperature relation  $\dot s/s = -3H$, where $s = 2\pi^2h_\mathrm{eff}T^3/45$ is the entropy density of the Universe and $H=(4\pi^3g_{\rm eff}/45)^{1/2}T^2/M_{P}$ and $g_\eff(T)$ and $h_\eff(T)$ are the effective number of relativistic energy and entropy degrees of freedom. This implies 
\begin{equation}
    \label{eq:time-temp-relation}
    \frac{\dot{T}}{T} 
    = 
    -\sqrt{\frac{4\pi^3}{45}}
	\frac{h_{\eff}(T)}{g_*^{^{1/2}}(T)}
	\frac{T^2}{M_{\textrm{P}}} \,,
\end{equation}
where
\begin{equation}
	\label{eq:gstr}
	g_*^{^{1/2}}(T) 
	\equiv 
	\frac{h_{\textrm{eff}}(T)}{g^{^{1/2}}_{\textrm{eff}}(T)}
	\left( 1 + \frac{T}{3h_{\textrm{eff}}} \frac{dh_{\textrm{eff}}}{dT}\right) \,.
\end{equation}
We want to replace also the momentum with a dimensionless variable. Using again the adiabaticity condition, one finds $(a/a_0)^3 = T^3_0h_\mathrm{eff}(T_0)/ T^3h_\mathrm{eff}(T)$, which suggests to define
\begin{equation}
\label{eq:dimless_mom}
    \xi \equiv \frac{k}{T_0a(t(T_0))} = \left(\frac{h_\mathrm{eff}(T_0)}{h_\mathrm{eff}(T)}\right)^{1/3}\frac{p}{T} \,,
\end{equation}
where $k$ is the comoving momentum, $p$ is the physical momentum, and $a_0$ is the scale factor evaluated at some reference temperature $T_0$, which we set equal to the reference mass: $T_0\equiv m_0$. 

The Boltzmann equation in dimensionless variables becomes
\begin{equation}
	\label{eq:comov_BE}
\frac{d f(\xi,x)}{d x}
	=
	\sqrt{\frac{45}{4\pi^3}}
	\frac{g_*^{^{1/2}}\big(\frac{m_0}{x}\big)}{h_{\eff}\big(\frac{m_0}{x}\big)}
	\frac{x M_{\textrm{P}}}{m_0^2}
	\sum \mathcal{C}_{\textrm{coll}}[f] \,.
\end{equation}
This Boltzmann equation, written in comoving variables, can be solved numerically by discretizing in variables $x$ and $\xi$. Given such a discretization, the role of the parameters $m_0$ and $T_0$ is to tune the dimensionless variables to probe the desired temperatures and physical momenta. Before describing this process in detail, we must first carefully describe the structure of the collision terms $\mathcal{C}_{\textrm{coll}}$. 
%
\subsection{Collision integral} 
%
The collision term for generic two-particle interactions $ 1 2 \leftrightarrow 3 4$ is given by
\begin{equation}
	\label{eq:C_coll}
	\mathcal{C}_{\textrm{coll}}[f_1]
	=
	\frac{1}{2E_1}\int d{\rm{PS}}_{234}
  \Lambda\left( f_1, f_2, f_3, f_4 \right)
	|{M_{12\rightarrow 34}}|^2,
\end{equation}
where the integration measure over the phase space is
\begin{equation}
  d{\rm{PS}}_{ijk}=(2\pi)^4\delta^{(4)}(p_1+p_2-p_3-p_4){\rm d}^3\tilde{p}_i{\rm d}^3\tilde{p}_j{\rm d}^3\tilde{p}_k,
\end{equation}
with ${\mathrm{d}}^3\!\tilde{p} = {\mathrm{d}}^3\!p/[(2\pi)^3 2E]$. We always assume that labels $i=1,2,3,4$ denote all internal degrees of freedom associated with a given distribution function $f_i$. The phase space factor $\Lambda(f_1,f_2,f_3,f_4)$ is defined as
\begin{align}
    \label{eq:phase_space_factor}
  \Lambda\left( f_1, f_2, f_3, f_4 \right) &=
  f_3f_4\left(1 \pm f_1 \right)\left(1 \pm f_2\right) 
  \nonumber \\
  &- f_1f_2\left(1 \pm f_3 \right)\left(1 \pm f_4 \right),
\end{align}
with $+\,(-)$ corresponding to the boson (fermion) case. Finally, $|M_{12\rightarrow 34}|^2$ is the matrix element squared, summed or integrated over the internal degrees of freedom associated with the labels $i=2,3,4$. The matrix element squared is also assumed to contain all relevant symmetry factors for the initial and final states. The collision integral $\mathcal{C}_{\textrm{coll}}$ naturally 
splits into the \textit{backward} and \textit{forward} terms, given by
\begin{equation}
	\label{eq:C_source}
	\mathcal{C}_{\mathrm{BW}}
	\equiv
	\frac{1}{2E_1}\int
	d{\rm{PS}}_{234}
	 f_3f_4\left(1 \pm f_1 \right)\left(1 \pm f_2\right)
	|M_{12\rightarrow 34}|^2
\end{equation}
and
\begin{equation}
	\label{eq:C_sink}
	\mathcal{C}_{\mathrm{FW}}
	\equiv
	-\frac{1}{2E_1}\int
	d{\rm{PS}}_{234}\, 
	f_1f_2\left(1 \pm f_3 \right)\left(1 \pm f_4 \right)
	|M_{12\rightarrow 34}|^2.
\end{equation}

The phase space integration of the collision integrals has been studied in the context of neutrino astrophysics for massless neutrinos in~\cite{Yueh:1976aa} and later for nonzero neutrino masses in~\cite{Hannestad:1995rs}. Similar methods were also developed, e.g.,~in~\cite{Semikoz:1995rd, Dolgov:1997mb, HahnWoernle:2009qn, Oldengott:2014qra, Hannestad:2015tea}. Here we follow the strategy of~\cite{Hannestad:1995rs} to reduce the fully general, initially nine-dimensional collision integrals down to four dimensions. The momentum dependence of the matrix elements prevents making further analytic simplifications.
 
Different from~\cite{Hannestad:1995rs}, we treat the forward and backward collision processes separately. This makes the numerical implementation more stable by avoiding the need to interpolate the unknown phase space distribution functions $f(p,t)$ in between the integration grid points. Full details of the reduction are given in Appendix~\ref{app:collisionterms}. The final result for the reduced backward term~\eqref{eq:C_source} is given by~\eqref{eq:BW_final} and for the forward term~\eqref{eq:C_sink}
by~\eqref{eq:FW_final}. 

\subsection{Discretization}

We solve Eq.~\eqref{eq:comov_BE} numerically by discretizing the momentum grid $\xi \rightarrow \xi_j$, $j = 1,\dots , N_\xi$, with uniform spacing in logarithmic scale. If the production processes spread over several orders of magnitude in temperature, this allows one to cover a sufficient range of momenta to reach the required accuracy.
After discretization, the Boltzmann equation \eqref{eq:comov_BE} becomes an initial value problem consisting of a coupled set of ordinary differential equations for $f_a(x,\xi_j) \equiv f_{aj}(x)$, over some temperature range $x$, that must be solved simultaneously\footnote{This makes the problem highly vectorizable but not easily parallelizable. We use \textsc{matlab} and, in particular, its stiff ode15s routine.} for each degree of freedom $a$ and the momentum mode $\xi_j$,
\begin{equation}
  \frac{d}{d x} f_{aj}(x)
     = \alpha(x) \sum\limits_\textrm{coll.} \mathcal{C}_{aj}(x).
\label{eq:disc_BE}
\end{equation}
Here $\alpha(x)$ is the prefactor given in the rhs of Eq.~\eqref{eq:comov_BE}, and the sum runs over all collision terms that contribute to evolution of $f_{aj}$. Here we separated the degrees of freedom (labeled by $a$) from the discretized momentum variable. Indeed, each different particle species, and each helicity or polarization state within a species, in general has its own independent unknown distribution function, which the collision terms couple with each other. Some hierarchies between the interaction rates may allow simplifying the equation network, such as helicity equilibrium due to rapid helicity flips. This can be easily incorporated by imposing the degeneracies and introducing the corresponding averaged matrix elements. We will typically assume that initially $f_{aj}=0$ for the dark sector particle distributions. This is justified when we start early enough in time, i.e., high enough temperature, and it allows us to track to which degree each species thermalizes before it decays or its distribution freezes.

As an example, on collision term discretization we show how the backward term~\eqref{eq:C_source} is implemented. Although the Boltzmann equation ~\eqref{eq:comov_BE} is solved in dimensionless momentum $\xi$, the collision term reduction in Appendix~\ref{app:collisionterms} is done in terms of the physical momentum $p$. The physical momentum $p_j$ corresponding to dimensionless momentum $\xi_j$ at a given temperature $T=m_0/x$ is then obtained by inverting Eq. \eqref{eq:dimless_mom},
\begin{equation}
    p_j(x) = \xi_j 
    \frac{m_0}{x}
    \left(\frac{h_\mathrm{eff}(\tfrac{m_0}{x})}{h_\mathrm{eff}(T_0)}\right)^{1/3} \,.
\end{equation}
Then backward term~\eqref{eq:C_source} can be reduced to~\eqref{eq:BW_final}, given in discretized form as
\begin{align}
	\mathcal{C}_{1j}^\mathrm{BW}(x) = 
    \frac{1}{16 E_{1j}}&
	\sum_k\sum_l 
	\frac{ \Delta p_k p_k^2 }{2\pi^2E_{3k}}
	\frac{ \Delta p_l p_l^2}{2\pi^2E_{4l}}
	\nonumber \\ 
	&\times F_{1jkl}(x)
	\, 
	\Lambda^\mathrm{BW}_{1jkl}(x) \,,
\end{align}
where the discretized backward phase space factor is
\begin{equation}
    \Lambda^\mathrm{BW}_{1jkl}(x) = f_{3k}(x) f_{4l}(x)\left[1\pm f_{1j}(x) \right]\left[1\pm f_{2jkl}(x) \right]
\end{equation}
and $F_{1jkl}(x)$ is the angular integral over the matrix element squared defined in Eq.~\eqref{eq:F_function}. Here the superscripts $1,\dots,4$ denote the particle species (in the sense described above) involved in the $12 \lra 34$ process, $f_{ai}(x)$ is the value of distribution function of the particle "$a$" with momentum $\xi_i$ at temperature $T=m_0/x$ and energy $E_{ai}=(p_i^2 + m^2_a)^{1/2}$. The label $j$ refers to the species 1, whose collision term we are computing and it is not summed over. The momentum space matrix structure $f_{2jkl}$ of species 2 follows from the four-momentum conservation. 

It is essential to note that the all matrices $F_{1jkl}(x)$ can be precalculated and replaced by numerical fit functions for all relevant processes  before solving the Boltzmann equations. This fitting procedure can be done very accurately and it is pivotal for the efficiency of the numerical code. The general flow of our implementation then is as follows:
\begin{itemize}
    \item[1)] Define theory, whose masses and couplings may depend on temperature.
    
    \item[2)] Create grids and define the entropy and energy degrees of freedom functions $h_\textrm{eff},g_\textrm{eff},g_*$.
    
    \item[3)] Determine relevant interactions and compute their matrix elements and cross sections.
    
    \item[4)] Precalculate the $F_{1ijk}(x_i)$ matrices for all species involved
    following Appendix~\ref{app:collisionterms}.
    
    \item[5)] Define the initial conditions for all distributions involved in the network, Eq.~\eqref{eq:disc_BE}.
    
    \item[6)] Solve the Boltzmann system numerically using a suitable ordinary differential equation solver.
    
\end{itemize}

This formulation is generic enough to allow for dynamical changes that modify the parameters of the theory during the evolution, such as phase transitions. In the next sections, we show in detail the results of this implementation in simple hidden sector models connected with the SM via the Higgs portal.

%
\section{First benchmark model: the singlet scalar extension} 
\label{sec:modelSS}
%

An extension of the SM by a real singlet scalar $S$, coupled with the SM Higgs doublet via the renormalizable operator $|H|^2S^2$, the "Higgs portal", provides a simple paradigm for a dark sector. Since its early introduction~\cite{Silveira:1985rk,McDonald:1993ex} this type of model building has started to gain more attention as benchmarks for experimental searches of particle dark matter~\cite{Burgess:2000yq,Barger:2007im,Farina:2009ez,Cline:2013gha}. Therefore, this model is a natural starting point for the tests of the computational method we have developed. Since the existing literature on  this model is large and its phenomenology has been thoroughly exposed already, our discussion here will be brief; we will introduce only the necessary formulas and focus on the comparison of our approach with other approximate computation schemes. The singlet model is defined by the Lagrangian
\begin{equation}
  {\cal L_{\rm SSM}} = \frac12 (\partial_\mu S)^2 - V(S,H) +{\cal L}_{\rm SM} \,,
\label{L_singlet}
\end{equation}
where the scalar potential is given by
\begin{align}
	\label{eq:scal_pot}
	V(S,H) =&
	 -\mu^2_\H |{H}|^2
	 -\frac{1}{2}\mu_\S^2 S^2
	 \nonumber\\
	&+ \lambda_\H |{H}|^4
	+ \frac{\lambda_\S}{4}S^4
	+ \frac{\lambda_{\HS}}{2} |{H}|^2 S^2 \,,
 \end{align}
and the gauge interactions of the Higgs doublet $H$ are contained in ${\cal{L}}_{\textrm{SM}}$. The stability of the potential requires that the quartic couplings $\lambda_\S$ and $\lambda_\H$ of the singlet and the Higgs fields are positive, but the portal coupling could be negative, as the stability of the potential requires just that $\lambda_\HS>-2\sqrt{\lambda_\S\lambda_\H}$. However, here we will only consider positive values of $\lambda_\HS$. 

If the singlet scalar mass $m_s$ is just below half of the Higgs mass $m_\H = 125.25 \pm 0.17$ GeV~\cite{ParticleDataGroup:2020ssz}, the Higgs mediated inelastic processes are resonantly enhanced but elastic processes maintaining the kinetic equilibrium are not, so one would expect the nonequilibrium effects to be relevant. Indeed, if the singlet $S$ is required to constitute all of the dark matter and remain compatible with the current direct detection experiments, its mass is constrained  to $m_s \in[56,62.5]$ GeV~\cite{Cline:2013gha}, where precision computations are required to address the dark matter phenomenology~\cite{Ala-Mattinen:2019mpa}. 

In our current, fully momentum-dependent setup, solving the singlet scalar relic density in this region entails solving the following Boltzmann equation:
\begin{equation}
    \partial_x f_s(\xi,x) 
    = \mathcal{C}^I_{ss \lra jj} + \mathcal{C}^I_{h \rightarrow ss} + \mathcal{C}^E_{sf \lra sf}\,,
\end{equation}
where $s$ refers to the singlet, $j=\{\tau,c,b,t,h,W,Z \}$, and $f=\{\tau,c,b\}$. In this case we only need a dynamical equation for $s$, since all SM particles can be assumed to be in thermal equilibrium. To correctly include kinematics when the Higgs mediator in $\mathcal{C}^I_{ss \lra jj}$ becomes on shell, we have included the on-shell Higgs decay $\mathcal{C}^I_{h \rightarrow ss}$ as a separate contribution and take all Higgs mediators in $\mathcal{C}^I_{ss \lra jj}$ to be off shell as described in Appendix~\ref{app:RIS}. Resonant inelastic processes deplete and overpopulate specific momentum states, which tends to bring the distribution function $f_s$ out of kinetic equilibrium. Elastic interactions, on the other hand, tend to restore the kinetic equilibrium, and if they are sufficiently fast, the standard thermal averaged treatment~\cite{Gondolo:1990dk,Griest:1990kh,Cline:2013gha} suffices. However, the Higgs resonance is particularly sharp,%
\footnote{$\Gamma_h \simeq 4$ MeV, so the width of the resonance in the $\sqrt{s}$ variable is $\sqrt{m_h\Gamma_h}\simeq 0.7$ GeV. } 
and one cannot \apriori assume that the elastic processes can maintain the kinetic equilibrium to high accuracy. The issue has already been studied using momentum-dependent methods~\cite{Binder:2017rgn,Ala-Mattinen:2019mpa}, however, using some approximations in the treatment of the collision integrals.

%
\begin{figure*}
\centering
\hspace{-1em}
\begin{minipage}[b]{.57\textwidth}
    \includegraphics[width=1\textwidth]{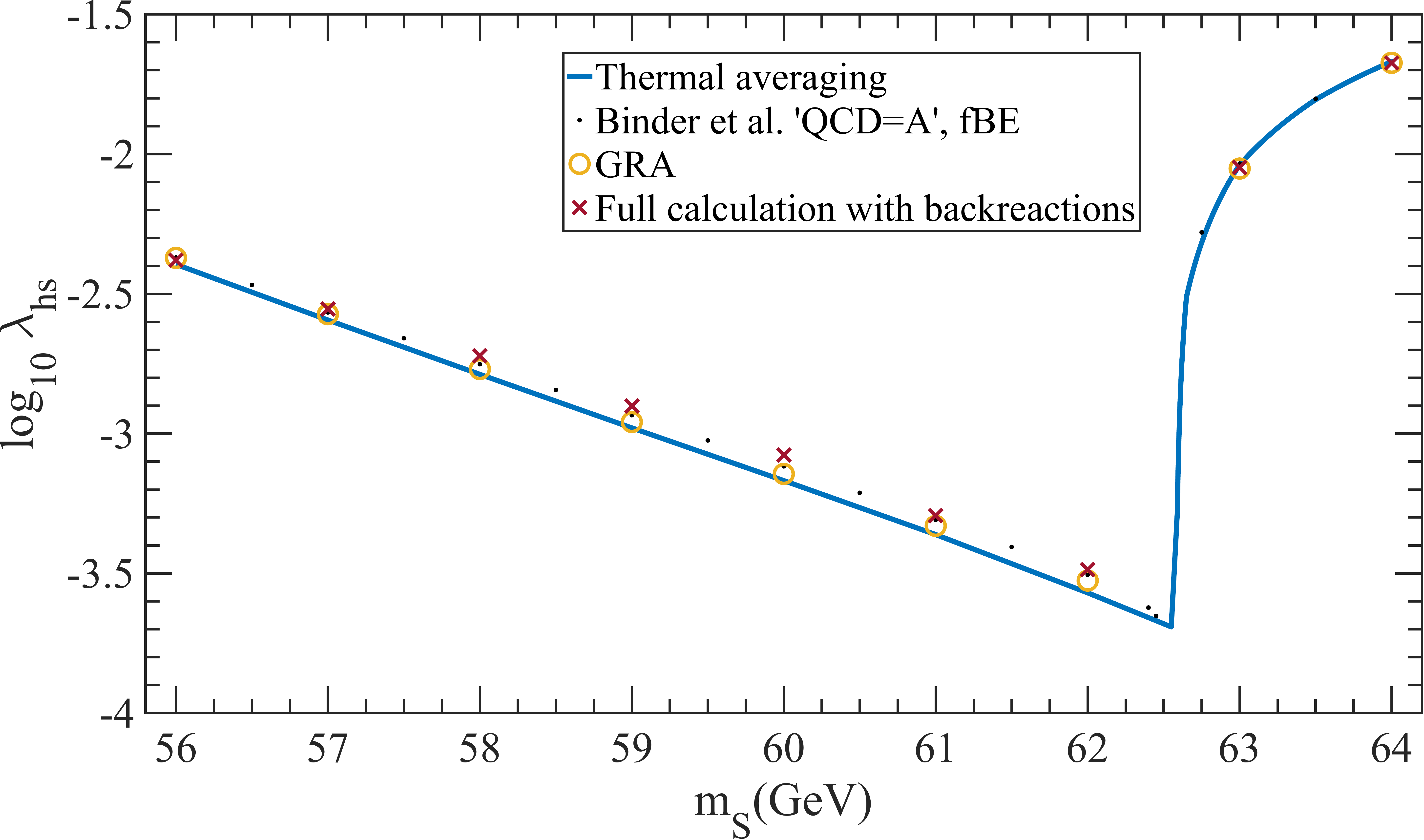}
    \end{minipage}\quad
\begin{minipage}[b]{.353\textwidth}
    \includegraphics[width=1\textwidth]{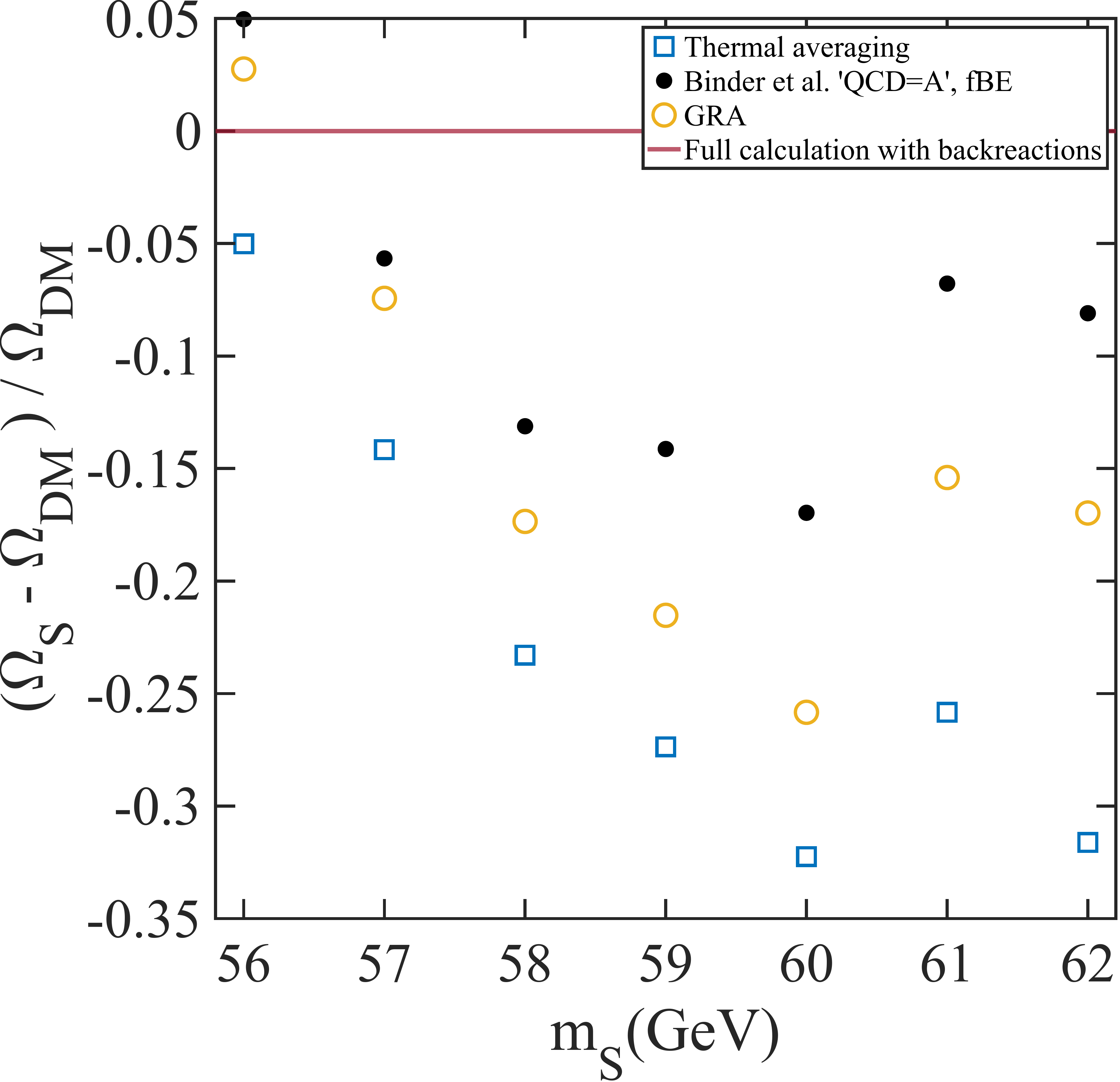}
\end{minipage}
\caption{ \textbf{Left:} fixed relic densities $\Omega_s h^2 = 0.1193$ from three different methods: momentum-independent thermal averaging (solid blue line), momentum-dependent generalized relaxation approximation~\cite{Ala-Mattinen:2019mpa} (yellow circles), and complete momentum-dependent computation with full backreaction terms (red crosses). Black dots are the full Boltzmann solution~\cite{Bringmann} corresponding to the method introduced in Ref.~\cite{Binder:2017rgn}, for the case "QCD=A". \textbf{Right:} relative difference in $\Omega h^2$ when $\lambda_\HS$ is fixed by the full calculation to yield $\Omega_s h^2 = 0.1193$ (red crosses on the left). Yellow circles correspond to relaxation approximation~\cite{Ala-Mattinen:2019mpa}, blue squares to the momentum-independent method, and black dots correspond to the results of Ref.~\cite{Binder:2017rgn}.}
\label{fig:higgs_resonance}
\end{figure*}
%

In left panel of Fig.~\ref{fig:higgs_resonance} we show the contours of $\Omega_s h^2 = 0.1193$ in the singlet mass and the portal coupling plane using different approximations. The results from the full computation implemented in this work are shown by the red crosses, while the standard thermal averaged result is shown by the solid blue line. The yellow circles correspond to the calculation using the momentum-dependent, generalized relaxation approximation (GRA) method of Ref.~\cite{Ala-Mattinen:2019mpa}, and finally, the calculation in Ref.~\cite{Binder:2017rgn}, using truncated expansions for the elastic collision integrals, is shown by the black dots. GRA calculation is similar to the one described in this paper, except for the numerically expensive elastic backward term $C_{sf \leftarrow sf}$, given by~\eqref{eq:C_source}. In the GRA method, this is treated in a simplifying approximation; for more details, see~\cite{Ala-Mattinen:2019mpa}.

\begin{figure*}
\centering
\begin{minipage}[b]{.3\textwidth}
    \includegraphics[width=\textwidth,keepaspectratio]{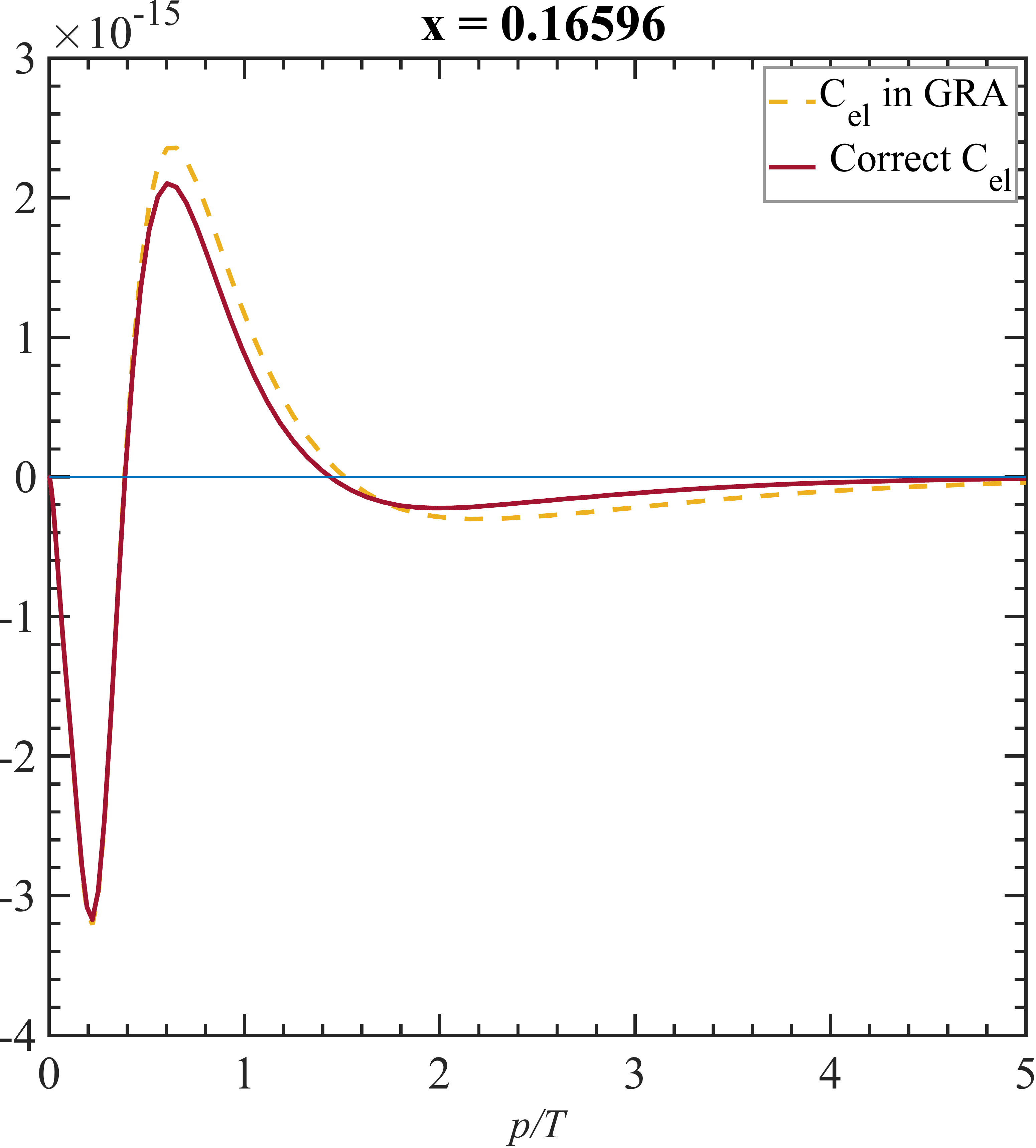} 
\end{minipage}\quad
\begin{minipage}[b]{.311\textwidth}
    \includegraphics[width=\textwidth,keepaspectratio]{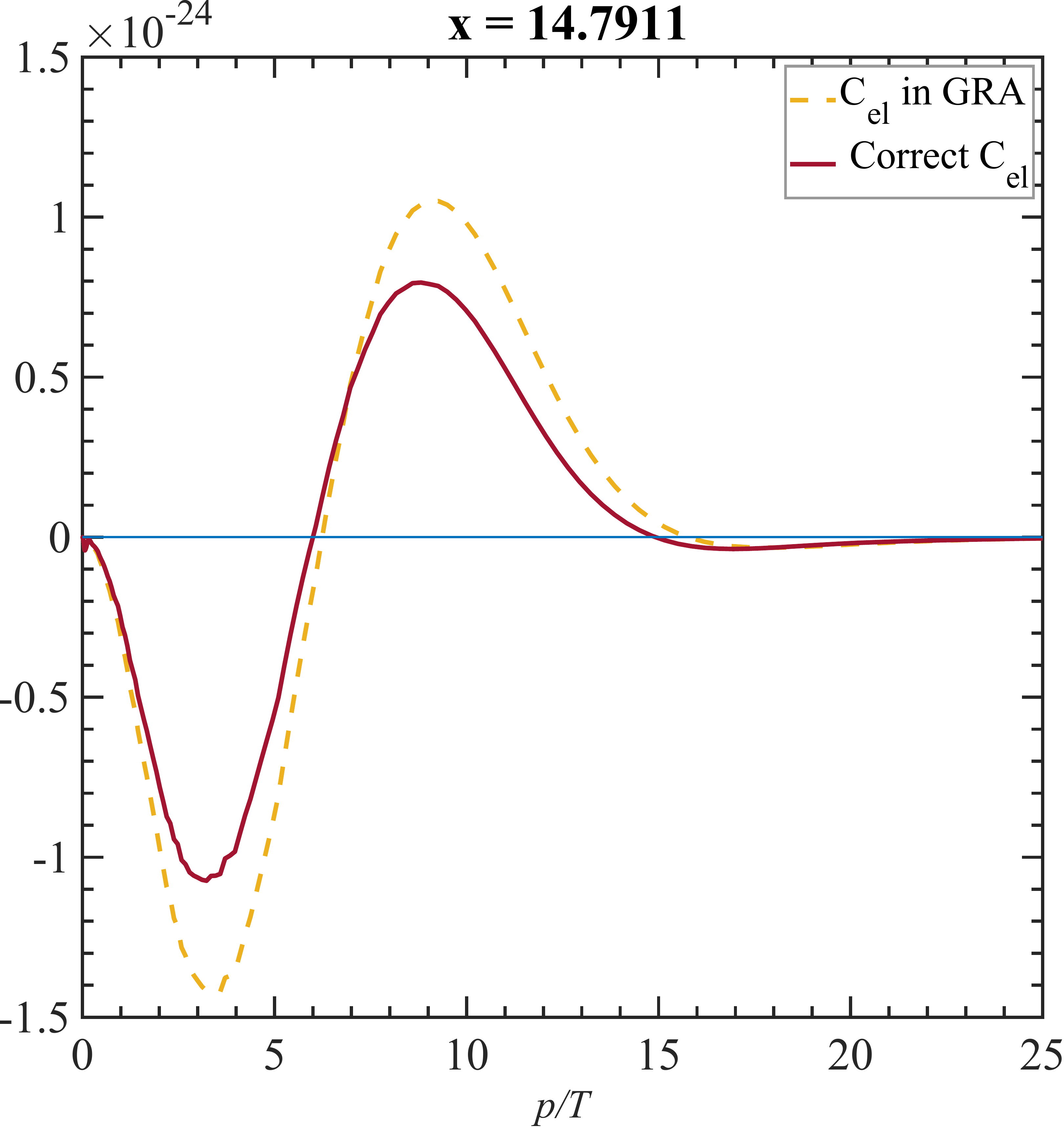}
\end{minipage}\quad
\begin{minipage}[b]{.3\textwidth}
    \includegraphics[width=\textwidth,keepaspectratio]{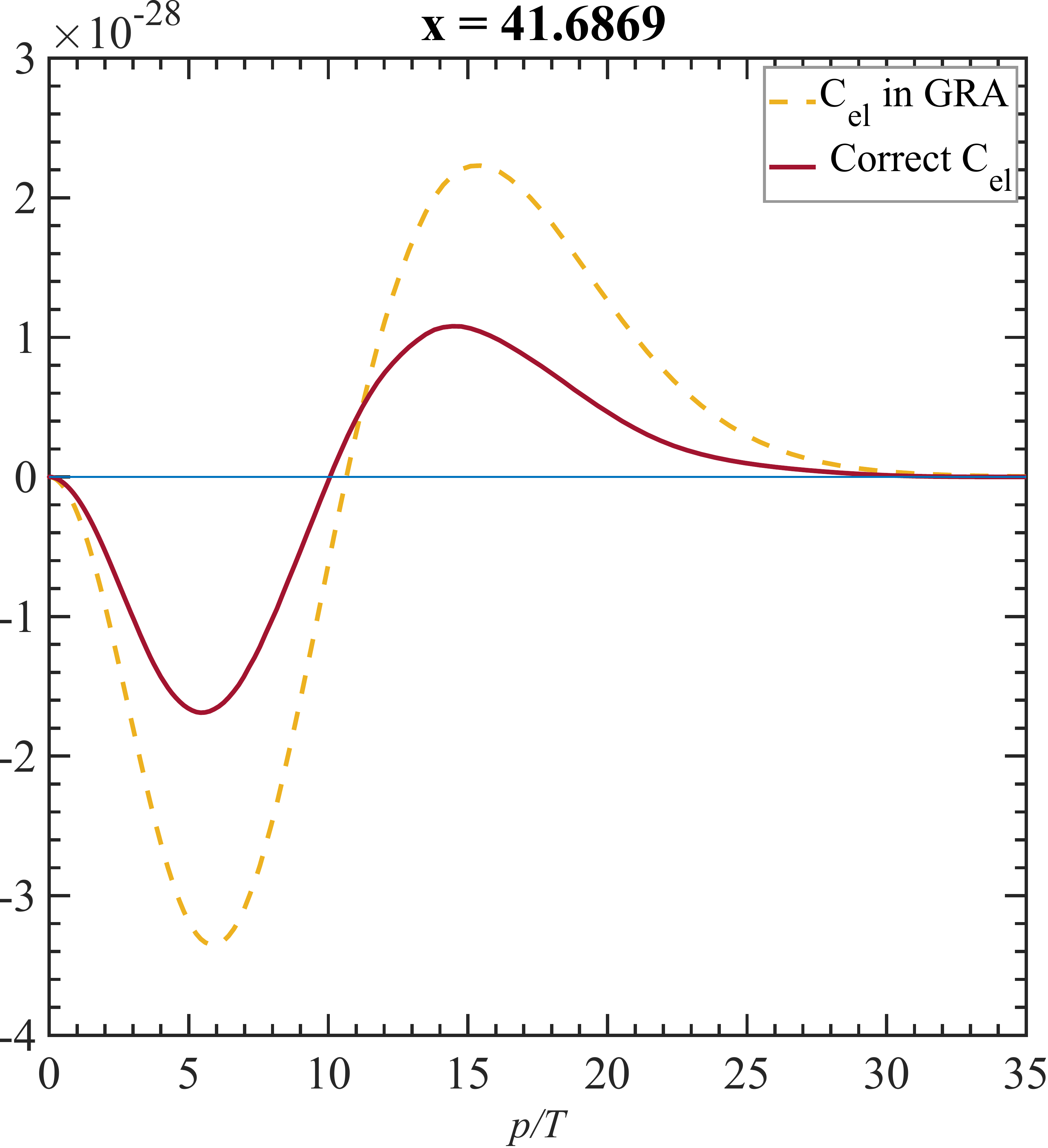}
\end{minipage}
\caption{Elastic integrals $(p^2/2\pi^2)C_{\rm el}(p)$ evaluated at three different temperatures $x = m_s/T$. The dashed yellow line shows the elastic integral resulting from generalized relaxation approximation used in~\cite{Ala-Mattinen:2019mpa}. The red solid line shows the correct elastic collision integral containing the full backward collision term given by Eq.~\eqref{eq:BW_final}.}
\label{fig:comp_ElasticBW}
\end{figure*}

When displayed in the logarithmic scale, all calculations appear to roughly agree. Plotting on linear scale (right panel in Fig.~\ref{fig:higgs_resonance}) reveals the significance of the deviations. The cause for the difference between the full and the GRA calculations is seen in Fig.~\ref{fig:comp_ElasticBW}, which shows the elastic collision integrals computed exactly (red solid line) and in the GRA (yellow dashed line). At high temperatures, the GRA method works well, but around the freeze-out temperature $x\sim 10$ it starts to overestimate the elastic integral that enforces the kinetic equilibrium. Eventually the error becomes of order $\sim 2$, but only well after the freeze-out $x\sim 40$. This tendency was already noted in~\cite{Ala-Mattinen:2019mpa}, and by construction the GRA scheme is not expected to work to a high precision for distributions that already are very close to thermal equilibrium. However, when one {\em is} close to equilibrium, the absolute magnitude of the error is already small, and GRA slightly improves on the thermal approximation.
The results using a truncated expansion for the elastic collision integral from~\cite{Binder:2017rgn} are roughly comparable with the GRA.

%
%

Overall, we confirm that the effect of kinetic decoupling in the singlet model is not as dramatic as one might have initially guessed. All methods agree in the absence of resonant enhancement as they should. Even in the resonant region, by far the most important effect is to use the thermally averaged annihilation rate in the ZOPLW equations, first pointed out in~\cite{Griest:1990kh}. The elastic interactions with the Standard Model particles are surprisingly efficient in keeping thermal equilibrium as observed in~\cite{Ala-Mattinen:2019mpa}. However, should a DM particle be identified in the resonance region, a precision calculation of its abundance requires a full momentum-dependent calculation with an exact collision integral.

Finally, let us note that the use of Maxwell-Boltzmann (MB) statistics in the \emph{inelastic} collision integral causes about a 10\% error~\cite{Dolgov:1992wf} in final abundance. This is the case for all results in Fig.~\ref{fig:higgs_resonance}. We have checked that the corresponding error in \emph{elastic} collision integrals only affects the final abundance by $\lesssim 1\%$.

To our knowledge, this is the first analysis of the singlet scalar in Higgs resonance with full elastic backreactions. This is undoubtedly due to the heavy numerical cost of the method. On average, we found that the computation times in this particular example (tested using a 6-core i7 laptop with 16 GB of RAM) scaled as follows in different approximations: thermal averaging runs took $\mathcal{O}(1)$ s, generalized relaxation approximation runs $\mathcal{O}(1-10)$ min, and the complete calculation with backreaction $\gsim 10$ h per $(m_s,\lambda_{hs})$ pair. Even faster and yet accurate methods of solving the ZOPLW equations exist~\cite{Cline:2013gha} for the use of large scale parameter scans. Taking this hierarchy into account is obviously of paramount importance when making a choice of what method to use for a given problem. 

\FloatBarrier

%
\section{Second benchmark model: singlet scalar and fermion extension}
\label{sec:modelSSN}
%

As a second example, we consider an extension of the SM by a singlet scalar ($S$) and a singlet Dirac fermion ($N$). The Lagrangian of the model is
\begin{equation}
	\label{eq:symm_lagrangian}
	\mathcal{L} =
	\mathcal{L}_{_{\mathrm{ SM} }}
	+
	i\overline{N}\slashed\partial N + \frac{1}{2}\left( \partial_\mu S\right)^2
	- V(S,H)
	+ yS\overline{N}N \,,
\end{equation}
where we denote the SM Higgs doublet by $H$. Its gauge interactions are contained in the SM Lagrangian ${\mathcal{L}}_{\mathrm{SM}}$, while the potential terms are contained in the extended scalar potential $V(S,H)$ given by Eq.~\eqref{eq:scal_pot}. In this model, the fermion $N$ is a phenomenologically interesting candidate for cold DM~\cite{Roland:2016gli,Boyarsky:2018tvu}. We are interested especially in the keV mass range, where the nonequilibrium dynamics can be relevant~\cite{Merle:2015oja,Konig:2016dzg}, and the resulting nonthermal momentum distribution of DM may affect  the formation of large scale structures. We focus on the question of whether highly nonthermal momentum distributions found in~\cite{Merle:2015oja,Konig:2016dzg} survive when all elastic processes are included in the analysis.

We will assume a mass hierarchy $m_N \ll m_S$, with  $m_N \sim \textrm{keV}$ and $m_S \sim \mathcal{O}(10 - 1000)$ GeV. The $N$-fermion mass gets a contribution from nonzero vacuum expectation value (VEV) of the singlet scalar $m_N = \mu_N + y\langle S \rangle$. As the VEV can be quite large, we need to assume the Yukawa coupling to be tiny, $y \ll 1$, to keep $m_N$ around keV scale. The vacuum structure is determined by the scalar sector of the theory.
The field $H$ is the usual weak doublet
\begin{equation}
	H =
	\begin{pmatrix}
		\phi^+ \\ \phi^0
	\end{pmatrix}
	\equiv
	\frac{1}{\sqrt{2}}
	\begin{pmatrix}
		\phi_1 + i\phi_2 \\
		\phi_3 + i\phi_4
	\end{pmatrix} \,,
\end{equation}
which has a VEV, denoted by $v$, along the neutral direction,
$\phi_3=v+\phi$. The VEV of the singlet field $S$ is denoted by $\langle S \rangle \equiv w$ and we write $S=w+\sigma$. Inserting these parametrizations into Eq~\eqref{eq:symm_lagrangian}, setting the field fluctuations to zero and extremizing the full scalar potential leads to
\begin{align}
	w\left(-\mu_\S^2  + \frac{1}{2}\lambda_\HS v^2  + \lambda_\S w^2\right) &= 0 \,,
	\label{eq:dVdS_vac}
	\\
	v\left(-\mu_\H^2  + \frac{1}{2}\lambda_\HS w^2  + \lambda_\H v^2\right) &= 0 \,.
	\label{eq:dVdphi3_vac}
\end{align}
We use these conditions to eliminate $\mu_\S^2$ and $\mu_\H^2$. This leads
to the mass matrix for neutral scalars $\sigma$ and $\phi$,
\begin{equation}
	\label{eq:M2}
	M^2 \equiv
	\begin{pmatrix}
		2\lambda_\H v^2 	&	\lambda_\HS v w		\\
		\lambda_\HS v w 	&	2\lambda_\S w^2		\\
	\end{pmatrix} \,,
\end{equation}
which is diagonalized by the transformation to the mass eigenbasis.
We denote the mass eigenstates by $h_1$ and $h_2$, so the explicit relation
is
\begin{equation}
	\begin{pmatrix}
		h_1 \\
		h_2
	\end{pmatrix}
	=
	\begin{pmatrix}
		\cos\theta & -\sin\theta \\
		\sin\theta & \cos\theta    
	\end{pmatrix} 
	\begin{pmatrix}
		\phi \\
		\sigma
	\end{pmatrix} .
\end{equation}
We identify $h_1$  with the SM Higgs field and $h_2$ is a heavier scalar. Consistency with LHC data on Higgs couplings then requires $\sin\theta \lsim 0.23$~\cite{CMS:2016dhk,ATLAS:2019nkf}. We therefore set $m_2>m_1=125.25$ GeV and consider the physical masses $m_1$ and $m_2$ to be input parameters. We then solve the couplings $\lambda_{\rm H}$ and $\lambda_{\rm S}$ and the mixing angle $\theta$ in terms of the physical masses,the vacuum expectation values and the portal coupling $\lambda_\mathrm{HS}$ as
\begin{align}
	\lambda_\H &= \frac{m_1^2 \cos^2\!\theta + m_2^2 \sin^2\!\theta}{2v^2} \,,  \label{eq:lambdaH}
	\\
	\lambda_\S &= \frac{m_2^2 \cos^2\!\theta + m_1^2 \sin^2\!\theta}{2w^2} \,, \label{eq:lambdaS}
	\\
    \sin(2\theta) &= \frac{2\lambda_\HS v w}{(m_2^2 - m_1^2)} \,.	\label{eq:sin2theta}
\end{align}
Requiring $\sin(2\theta)$ to be positive
implies that $0\leq \lambda_\HS \leq \frac{m_2^2-m_1^2}{2vw} \equiv \lambda_\HS^{\textrm{max}}$.

The Feynman rules following from the Lagrangian \eqref{eq:symm_lagrangian} are tabulated in Appendix~\ref{app:feynmanrules}. In the special case of $w=0$, the treatment is more straightforward, as the fields $\sigma$ and $\phi$ are directly the mass eigenstates of the mass matrix. Without going into further details, we simply note that the Feynman rules of Appendix~\ref{app:feynmanrules} can be directly applied also in this case by letting $\theta,w \rightarrow 0$. In the limit of vanishing Yukawa coupling and singlet scalar VEV, $y,w \rightarrow 0$, the model reduces to the singlet scalar model from previous section.

With a slight abuse of notation, we denote the mass eigenstates by $\phi$ and $\sigma$, as this allows us to include the cases $w\neq 0$ and $w=0$ simultaneously. Then we can summarize the above construction as follows:  we have taken the masses $m_1\equiv m_\phi$, $m_2\equiv m_\sigma$, the portal coupling $\lambda_\HS$, and the vacuum expectation values $w$ and $v$ as the input parameters, and express other Lagrangian parameters in the scalar sector in terms of these. Furthermore, we fix $v=246$ GeV. Thus, the free parameters in this theory are $\{m_\sigma, \lambda_\HS, w, y,m_N \}$.

\subsection{DM production processes and coupled Boltzmann system}

The Lagrangian \eqref{eq:symm_lagrangian} allows for various production processes for the $N$ and $\sigma$. Processes of order $\mathcal{O}(y^2)$ are negligible and the relevant contributions under our assumptions are summarized in Table \ref{table:processes}. Because of nonzero Yukawa coupling and assumed mass hierarchy, eventually all produced $\sigma$ scalars will decay into $N$ fermions, which remains as a stable relic. Production of $\sigma$ scalars is determined by the portal coupling $\lambda_\HS$. Direct production of $N$ fermions from a SM heat bath is allowed by a nonzero mixing angle $\sin\theta$ between the scalars, but remains subdominant for the allowed small mixing angles. Therefore, the production of $N$ fermions proceeds mostly via $\sigma$-scalar decays, which is itself produced from a SM heat bath and whose number density can freeze (either via freeze-in or freeze-out mechanism) before it fully decays.
\begin{table}
\begin{tabular}{ c c } 
 Always open \qquad & \qquad  Open if $w>0$ \\
 \hline
 $\sigma\sigma \lra \phi\phi$       & $\phi \ra NN$             \\
 $\sigma\sigma \lra VV$             & $N\sigma \lra N\sigma$    \\ 
 $\sigma\sigma \lra ff$             & $N\phi \lra N\phi$        \\ 
 $\phi \ra \sigma\sigma$            & $Nf \lra Nf$              \\ 
 $\sigma \ra NN$                    &   \\
 $\sigma f \lra \sigma f$           &   \\
 $\sigma\sigma \lra \sigma\sigma$   &   \\
 $N\sigma \lra N\phi$               &   \\
\end{tabular}
\caption{Relevant tree-level production processes for singlet fermion $N$ and singlet scalar $\sigma$ in the model \eqref{eq:symm_lagrangian} assuming $y\ll 1$. Each $s$, $t$, and $u$ channel reaction can be mediated via both scalar fields: singlet $\sigma$ and SM Higgs $\phi$. Here $V=\{W,Z\}$ labels vector bosons and $f=\{\tau,c,b,t\}$ labels SM fermions. }
\label{table:processes}
\end{table}

To obtain the momentum distribution function for $\sigma$ scalar and $N$ fermion we must solve the following set of coupled Boltzmann equations:
\begin{align}
    \partial_x f_\sigma(\xi,x) 
       =& \; \mathcal{C}^I_{\sigma\sigma \lra jj} + 
        \mathcal{C}^I_{\phi \rightarrow \sigma\sigma} + 
        \mathcal{C}^I_{\sigma \rightarrow \N\N} 
   \nonumber \\  
       +& \; \mathcal{C}^I_{\N \sigma \lra \N \phi } +
        \mathcal{C}^E_{\sigma f \lra \sigma f } +
        \mathcal{C}^E_{\N \sigma \lra \N \sigma } 
        \,,
    \vspace{2mm}
    \\
    \vspace{2mm}
    \partial_x f_\N(\xi,x) 
    =& \; \mathcal{C}^I_{\sigma \lra \N\N} + 
      \mathcal{C}^I_{\phi \lra \N\N} + 
      \mathcal{C}^E_{\N \sigma \lra \N \phi } 
   \nonumber \\ 
     +&\; \mathcal{C}^E_{\N \sigma \lra \N \sigma } + 
        \mathcal{C}^E_{\N \phi \lra \N \phi } +
        \mathcal{C}^E_{\N f \lra \N f } 
        \,,
\label{eq:SSN_BE}
\end{align}
where again the SM states are denoted as $j=\{\tau,c,b,t,\phi,W,Z \}$ and $f=\{\tau,c,b\}$. The form of this equation shows one obvious fact about solving the momentum-dependent kinetic equations: most of the work involved goes to definition and computation of the various collision integrals. The $\phi$ and $\sigma$ propagators in the inelastic $2\rightarrow 2$ collision integrals $\mathcal{C}^I_{\sigma\sigma \lra jj}$ are taken to be off shell, as the on-shell contributions are already included separately in $\mathcal{C}^I_{\phi \rightarrow \sigma\sigma}$ and $\mathcal{C}^I_{\sigma \rightarrow \N\N}$. There are several different suggestions in literature as to how this real intermediate state (RIS) subtraction should be done, e.g.~\cite{Kolb:1979qa,Giudice:2003jh}. Here we are following the treatment of~\cite{Cline:1993bd,Cline:2017qpe}; see Appendix~\ref{app:RIS} for more details and discussion.

Different from previous treatments, we have also accounted for the three- and four-body final states from virtual boson decays using methods described in~\cite{Cline:2013gha}, as well as the one-loop corrections for quarks in the $\sigma\sigma \lra jj$ channel. Accounting for virtual boson decays and QCD one-loop corrections describe the SM states more accurately and slightly increase the $\mathcal{C}^I_{\sigma\sigma \lra jj}$ contributions in Eq.~\eqref{eq:SSN_BE}. This is good to keep in mind when comparing our results to, e.g., Ref.~\cite{Konig:2016dzg}, as in the case of $\sigma$ freezing out this slight increase causes the $\sigma$ to follow the SM heat bath a bit longer and slightly suppresses the final fermion distribution.

\subsection{Results and discussion when \texorpdfstring{$w=0$}{w=0}}

We first set the VEV of the singlet scalar to zero, so that the scalars do not mix. This leaves us with processes on the left column of Table~\ref{table:processes}. This setting is equivalent to the one studied in Ref.~\cite{Konig:2016dzg}, except that we have included the elastic processes $\sigma f \lra \sigma f$, $\sigma\sigma \lra \sigma\sigma$, and $N \sigma \lra N\phi$, which tend to suppress the nonthermal component in the momentum distribution of $N$ fermion. The two first processes can also lengthen the freeze-out time of $\sigma$ field, thus allowing it to be Boltzmann suppressed more before it freezes out and decays, which can reduce the late time production of $N$ fermions. The two-peaked nonthermal momentum distribution found in~\cite{Konig:2016dzg} results from $N$ being produced at two separate temperature scales (see~\cite{Dienes:2020bmn} for a comprehensive study). Hence, reducing the production at either temperature scale could prevent the momentum distribution from forming the double peak structure. The last two processes tend to restore the kinetic equilibrium by reducing the nonthermal component momentum distribution. In practice, we find their effect to be negligible.

%
\begin{figure}[t]
\includegraphics[trim=35mm 0mm 64mm 18mm, clip, width=0.47\textwidth]{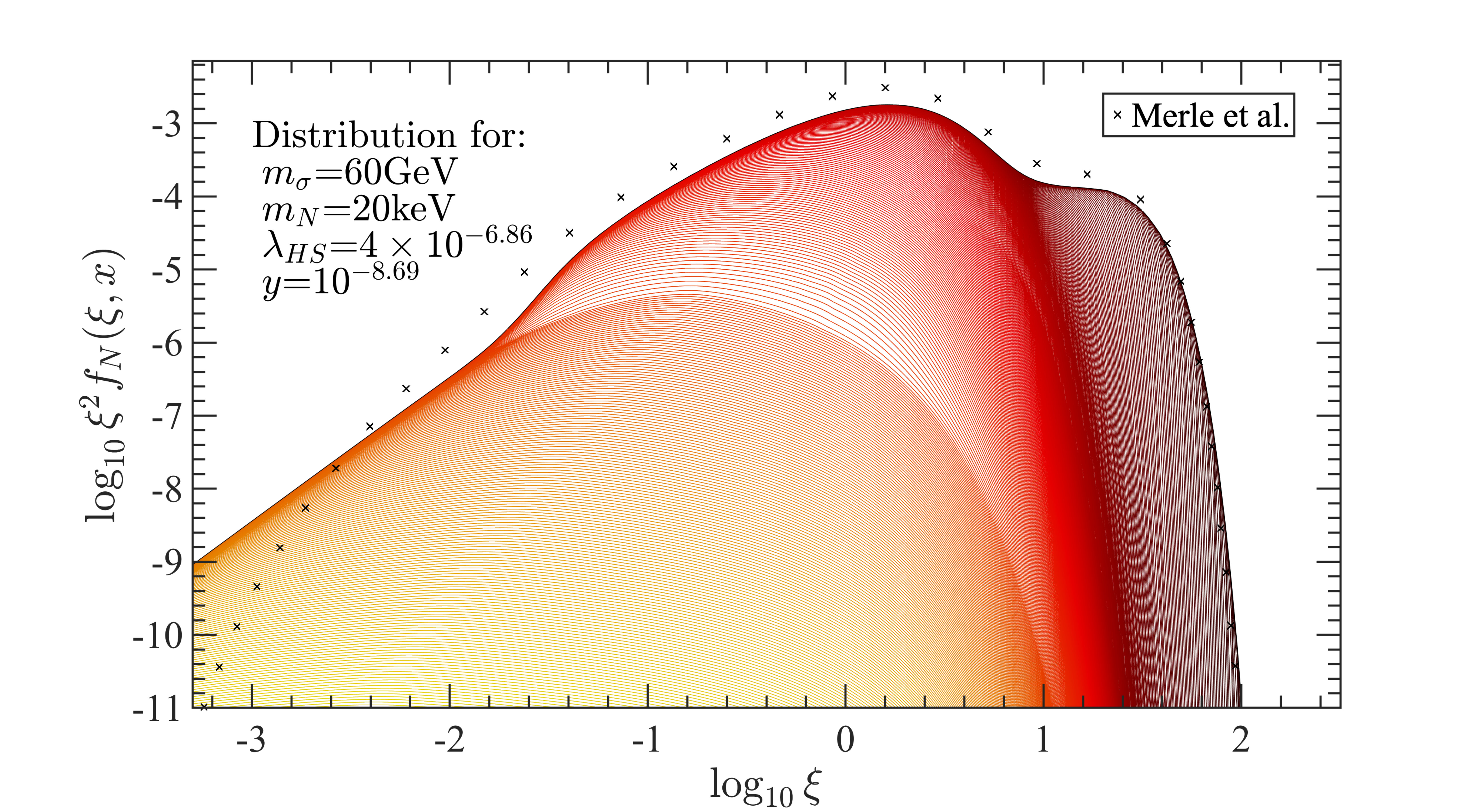}
\caption{Momentum distribution functions of singlet fermion obtained from our implementation versus the results in Fig. 5 of~\cite{Konig:2016dzg} (black crosses). Note that our convention to $\lambda_\HS$ differs from~\cite{Konig:2016dzg} by a factor of 4. Darker colors refer to later times and the black solid curve is the final frozen-in form. The low-end momentum tail is produced at temperatures much higher than the electroweak phase transition temperature, $T\gg T_\textrm{EWPT} = 150$ GeV. The temperature grid range is $\log_{10}(x) \in [-3.5, 3]$ and we used $m_0=T_0=m_h$ as the reference scales in~\eqref{eq:dimless_mom} and \eqref{eq:comov_BE}.  }
\label{fig:SSN_yield_and_distr}
\end{figure}
%

Figure~\ref{fig:SSN_yield_and_distr} shows our solution for the momentum distribution function of the fermion $N$ for parameter values  $\{m_\sigma =60\ \textrm{GeV},\ m_N =20\ \textrm{keV},\ \lambda_\HS = 4\times 10^{-6.86},\ y = 10^{-8.69},\ w=0\}$. These values are chosen to correspond to the case presented on the right-hand panels of Fig. 5 in Ref.~\cite{Konig:2016dzg} and we have included their final result as black crosses in our Fig.~\ref{fig:SSN_yield_and_distr}. Clearly the elastic processes are not sufficiently fast to suppress the formation of the nonthermal momentum distribution. The reason is quite clear {\em a posteriori}: The second peak at high momenta forms only when $\sigma$ production from the SM heat bath has already stopped and the remaining scalars decay into $N$ fermions. At this point, the SM Higgs is also heavily Boltzmann suppressed. The elastic channel $N\sigma \lra N\phi$, which would most effectively thermalize the momentum distribution of $N$ fermions, is therefore suppressed due to absence of both scalars in the thermal bath.

We have also checked the other light ($m_\sigma < m_h/2$) and heavy ($m_\sigma > m_h/2$) scalar (corresponding to feebly and weakly interacting massive particles) scenarios as discussed in~\cite{Konig:2016dzg} and found that our results agree with theirs to the same extent as in the above example. In summary, we find that the model presented in~\cite{Konig:2016dzg} is indeed inherently nonthermal in parts of its parameter space when assumptions $y\ll 1$  and $w=0$ hold. Our numerical results for the distribution $f_N(\xi,x)$ differ noticeably from~\cite{Konig:2016dzg}, but the difference does not come from elastic interactions, but from our more accurate evaluation of the inelastic interaction rates.

\subsection{Results and discussion when \texorpdfstring{$w>0$}{w>0}}

We now let the singlet obtain a nonzero VEV $w>0$. The nonzero mixing angle induced by the broken symmetry is given by Eq.~\eqref{eq:sin2theta}. It increases the $\sigma$ and $N$ particle production rates and opens new elastic channels for the $N$ field. All relevant reaction channels are given in Table~\ref{table:processes}, where also the channels on the right column now contribute. 

\begin{figure}[t]
\hspace{-1em}
\includegraphics[trim=35mm 0mm 64mm 20mm,clip,
width=0.47\textwidth]{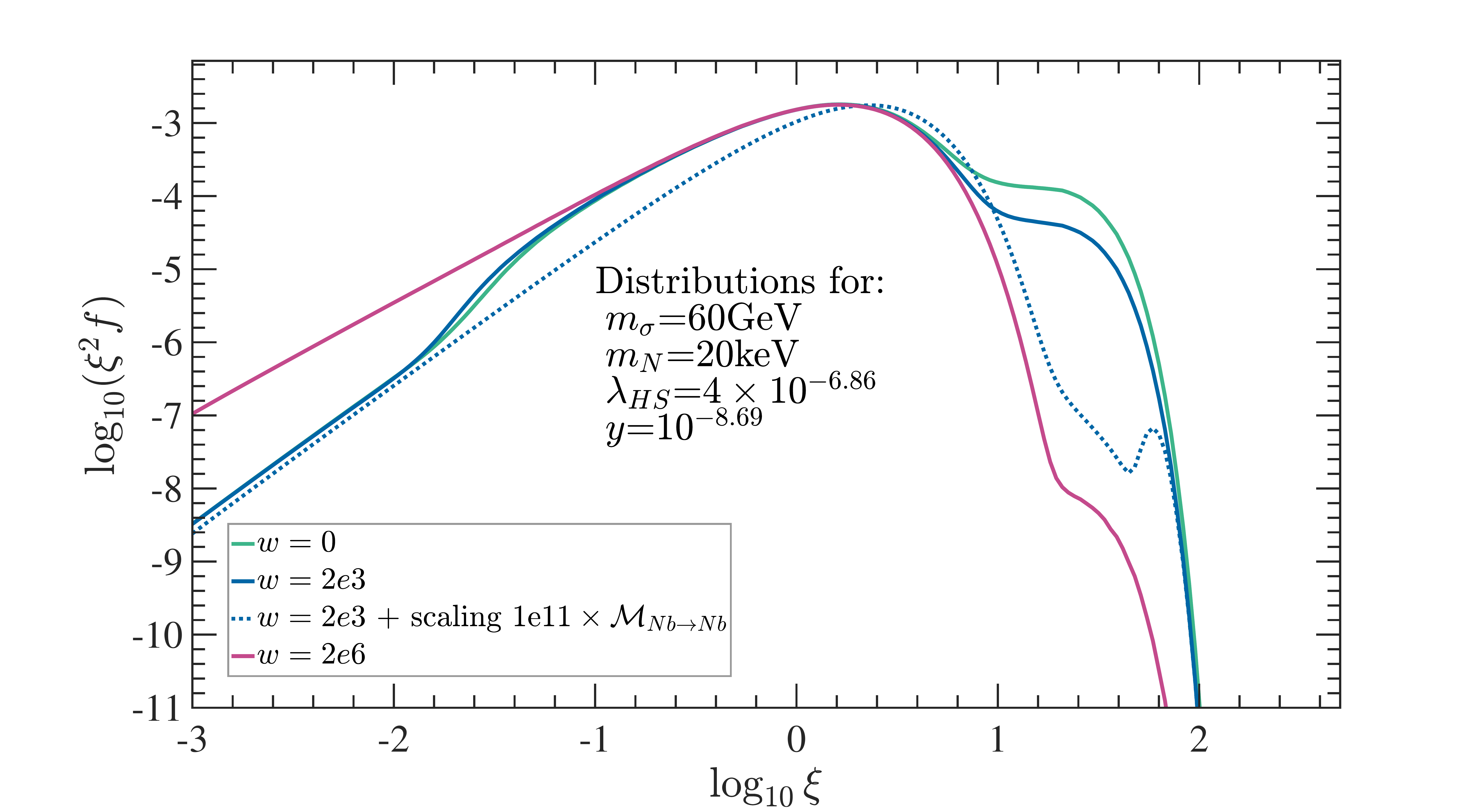}
\caption{Frozen-in momentum distributions for the fermion $N$ for varying $w$. The green solid line corresponds to the case in Fig.~\ref{fig:SSN_yield_and_distr}, and the blue dotted line shows an example of forced thermalization when the elastic rates are increased by hand, which we show only as a proof of concept to show that the kinematics work as indented and the system then tends toward thermal distribution.}
\label{fig:distr_comp_w>0}
\end{figure}

\begin{figure}[b]
\hspace{-1em}
\includegraphics[trim=27mm 0mm 12mm 5mm, clip ,width=0.485\textwidth]{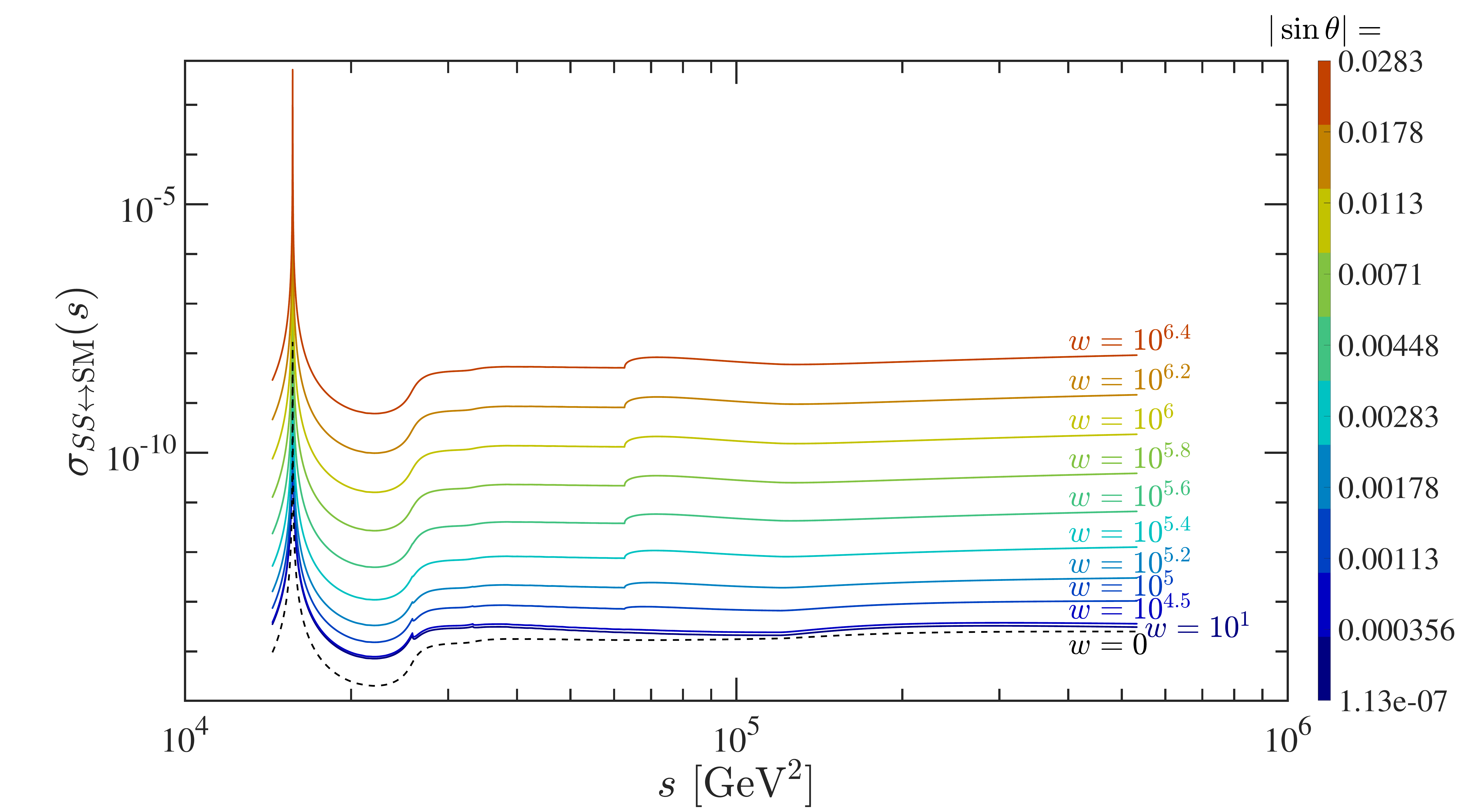}
\caption{Cross section between the singlet scalars $S=w+\sigma$ and SM particles plotted over a representative range of incoming energies with varying singlet VEV $w$ (mixing angle $|\sin\theta|$). Other parameters are $\{m_S=60\,\mathrm{GeV},\,\lambda_{HS}=4\times 10^{-6.86}\}$.}
\label{fig:sigma_ss-sm}
\end{figure}

\begin{figure}[t]
\hspace{-1.2em}
\includegraphics[trim=35mm 0mm 60mm 20mm, clip,width=0.47\textwidth]{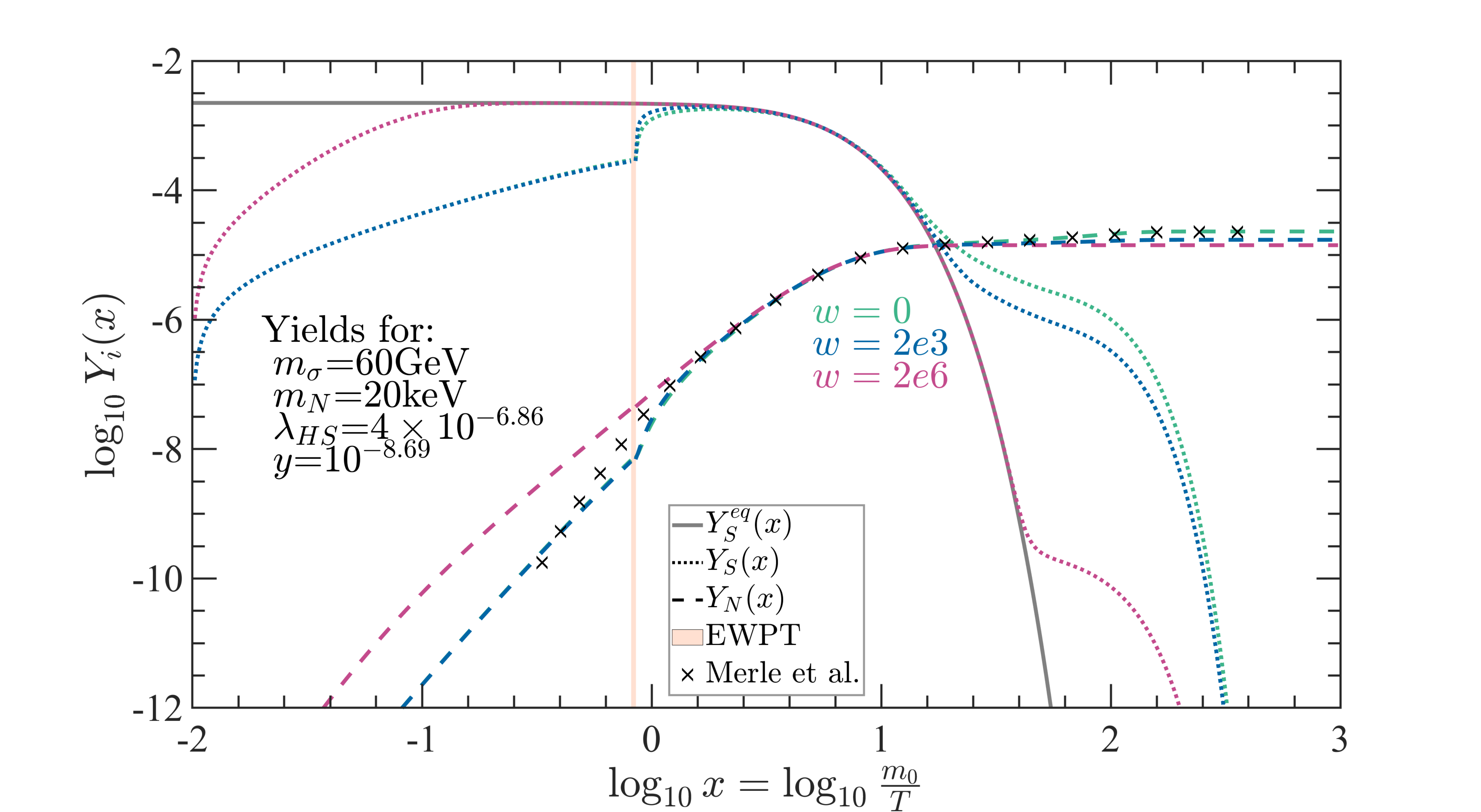}
\caption{Yields $\smash{Y\equiv \frac{n_i(x)}{s(x)}}$ for singlet scalar $S$ (dotted lines) and fermion $N$ (dashed lines). $m_0 = m_\H$ and colors show results from different choices of the singlet VEV $w=\{0,\,2\times 10^3,\,2\times 10^6\}$ GeV. Crosses show data from Fig. 5 of~\cite{Konig:2016dzg} corresponding to $w=0$ case.}
\label{fig:yield_comp_w>0}
\end{figure}

In Fig.~\ref{fig:distr_comp_w>0} we compare our results in the $w>0$ case with the previous $w=0$ case (shown with the green curve) to see the effect of additional channels and the increased interaction rates on the nonthermal double peak structure of the momentum distribution of $N$ fermions. We have chosen conservative values for the singlet VEV: $w=2\times 10^3$ GeV (blue curve) and $w=2\times 10^6$ GeV (red curve). In both, cases we find a strong suppression on the amplitude of the nonthermal second peak, which strongly increases for increasing $w$. However, this is again not due to elastic processes being effective in redistributing the quanta, but due to the fact that the inelastic interactions between the $\sigma$ scalar and SM heat bath strongly increase for increasing values of $w$, as illustrated in Fig.~\ref{fig:sigma_ss-sm}. As a result, the number density of $\sigma$ particles becomes more and more Boltzmann suppressed before they finally decay to $N$ fermions. This then suppresses or removes the high-momentum out-of-equilibrium tail from the $N$ distribution. The mechanism is clearly illustrated in Fig.~\ref{fig:yield_comp_w>0}, where we show the integrated yields of the singlet scalar fields (dotted lines) as a function of $m_0/T$ for the same choices of parameters as in Fig.~\ref{fig:distr_comp_w>0}. The effect on integrated neutrino distributions (dashed lines) is much smaller than to their momentum dependence, but it shows the right tendency as function of the $\sigma$-field abundance.

%
%

%
%

We found that the elastic interactions had negligible effect on the $N$ distribution in comparison with the Boltzmann suppression discussed above. Increasing the VEV of the singlet does result in stronger elastic rates for the fermion $N$, but they still fail to restore the thermal equilibrium. This is partly because the having $w>0$ also increases the overall production of $N$ fermions via inelastic and decay channels, which is the root cause for the nonthermal distributions. We then find that even in the $w>0$ case the elastic processes are never sufficiently fast to thermalize the fermion momentum distribution and the model~\eqref{eq:symm_lagrangian} is inherently nonthermal when $y\ll1$ and $m_N \ll m_S$. However, if a phase transition takes place in the singlet sector before the singlet field freezes out, the predictions for the amplitude of the out-of-equilibrium component in $N$ distribution change dramatically for a given set of parameters, in comparison to the $w=0$ case studied in~\cite{Konig:2016dzg}. However, this change is not due to increased elastic interactions, but due to change in inelastic rates, and to discover this effect the high-accuracy Boltzmann codes developed in this work are, in fact not necessary.

%
%

The inefficacy of the elastic rates in the examples studied above results from the smallness of the adopted values for the $y$ coupling and this does not imply that  elastic interactions were irrelevant in general. Instead of constructing a model just to make the elastic rates important, as a proof of concept, we simply increase the elastic process $Nf\lra Nf$ by hand to see how large it must be for a significant thermalization by elastic channels to take place. For $\{m_\sigma=60\ \textrm{GeV},\ m_N=20\ \textrm{keV},\ \lambda_\HS=4\times10^{-6.86},\  y=10^{-8.69},\ w=2\times 10^3\}$ we scaled the matrix element ${\mathcal{M}}_{Nf\ra Nf} \ra 10^{11}\times {\mathcal{M}}_{Nf\ra Nf}$. The nonscaled result is shown as a solid blue curve and the result after scaling as a dotted blue curve in Fig.~\ref{fig:distr_comp_w>0}. Thus, in this case, at the level of cross sections, the elastic rates are roughly $\sim 10^{-22}$ times too small to significantly thermalize the system.

\begin{figure}[t]
\hspace{-1em}
\includegraphics[trim=36mm 0mm 62mm 25mm, clip ,width=0.47\textwidth]{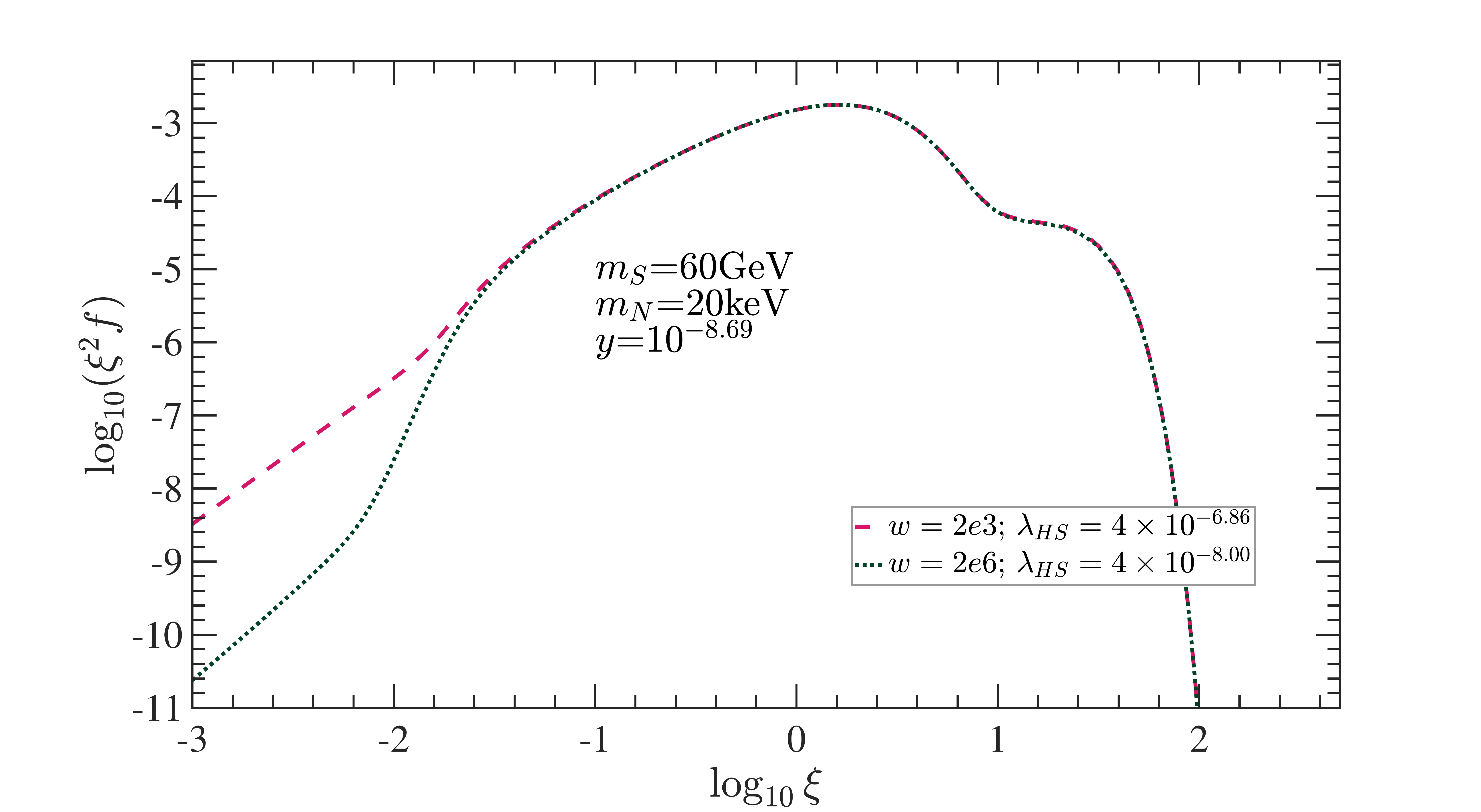}
\caption{ The final frozen-in momentum distribution of the fermion $N$ depends on the parameters $\{w,\lambda_{HS}\}$ so that increasing one can be countered by decreasing the other.}
\label{fig:lhs_scaling}
\end{figure}

Let us finally note that, while the introduction of the phase transition changes the predictions significantly for a given portal coupling, the same out-of-equilibrium distribution can be obtained in the spontaneously broken case for a different portal coupling. That is, there are degenerate subspaces in the $(\lambda_\HS, y;w)$ parameter space, where the same nonthermal momentum distribution can be obtained. In particular for a fixed Yukawa $y$, essentially only the interaction rate between the singlet scalar and SM heat bath is important in determining the degree to which the momentum distribution becomes nonthermal. As this rate is determined by a combination of $\lambda_\HS$ and $w$, we can always find such a $(w',\lambda'_\HS)$ pair that the original nonthermal behavior for $(\lambda_\HS, y;w=0)$ parameters is obtained with $(\lambda'_\HS, y;w'>0)$. This is shown in Fig.~\ref{fig:lhs_scaling}, where we plot the distributions with $y=10^{-8.69}$ for the cases $\{ w=2\times10^3\ {\rm GeV},\   \lambda_\mathrm{HS}=4\times 10^{-6.86} \}$ and $\{ w=2\times10^6\ {\rm GeV},\  \lambda_\mathrm{HS}=4\times 10^{-8}\}$, respectively. In each case we find exactly the same momentum distribution, apart from small deviations in the very small momenta. 

%
\FloatBarrier
%

%
\section{Conclusions}
\label{sec:checkout}
%

We have presented a computational method that is generally applicable for solving the coupled set of Boltzmann equations for phase space distribution functions of cosmic relics. Similar techniques have been developed earlier in the context of the neutrino astrophysics~\cite{Kainulainen:2001cb,Ghiglieri:2015jua,Hannestad:2015tea,Bodeker:2020hbo}, but until now they have not been carefully adapted to be used in the dark matter abundance calculations. For earlier implementations that use different levels of approximations for the collision integrals, see~\cite{Binder:2017rgn,Ala-Mattinen:2019mpa}. One of the main results is the compilation of these methods into a simple and generic form that provides a suitable starting point for their wider utilization in cases, where the standard thermal averaging methods fail and more accurate predictions in DM models are needed. 

We demonstrated our method in two models known to be sensitive to the momentum dependency of the phase space distribution.  Our first example concerned the freeze-out of a singlet scalar dark matter coupled to SM via the Higgs portal near Higgs resonance. In this case, the kinetic equilibrium approximation required by the usual ZOPLW equation cannot be assumed {\em a priori}. We performed a full momentum calculations, comparing our results to the ZOPLW approximation and two earlier momentum-dependent calculations, where further approximations of the form of the elastic collision integrals were made. 

We found that the approximation methods of Refs.~\cite{Ala-Mattinen:2019mpa,Binder:2017rgn} are in good agreement with our full results. Both methods present some improvement over the ZOPLW results even very near equilibrium, although they both slightly (the former a little more) overestimate the elastic rates.

As another example, we considered the model presented in~\cite{Merle:2015oja,Konig:2016dzg}. Using our methodology we extended their calculations to include the elastic processes and possible mixing between the two scalar states as a result of a phase transition in the singlet sector. We found that this could significantly alter the predicted size of the nonequilibrium distribution, the more so the larger the VEV of the singlet field. However, this difference was mainly due to changes in the inelastic rates, caused by the phase transition. Elastic rates turned out to be  inefficient and even when they were included, the momentum distribution of the singlet fermion remains inherently nonthermal. Overall, we find that, in a vast majority of cases, the momentum averaged methods work surprisingly well. 
\vspace{-0.3truecm}
%
\section*{Acknowledgments} 
%

We thank T.~Bringmann for correspondence and sharing data related to Fig. 1.
This work was supported by the Academy of Finland Grants No. 310130, No. 342777, and No. 318319. K. A-M. is funded by the doctoral program in Particle Physics and Universe Sciences in University of Helsinki.

%
\section*{Note added}
%

Recently, a paper presenting similar methods, also based on~\cite{Hannestad:1995rs}, appeared in~\cite{Du:2021jcj}. Our results agree qualitatively with theirs.

\appendix

%
\section{COLLISION TERM KINEMATICS} 
\label{app:collisionterms}
%

Here we give details of the integration of collision terms following closely the treatment of Ref.~\cite{Hannestad:1995rs}. Our goal is to clearly isolate the general structures in the collision terms that can be preevaluated prior to the solution of the Boltzmann equation.

\subsection{2-2 Scatterings} 

\noindent{\it Backward term.}  Different from~\cite{Hannestad:1995rs}, we treat the forward and backward collision integrals separately. We start from the backward term. In Eq.~\eqref{eq:C_source}, we use the four-momentum conservation to integrate over $\mathrm{d}^3\! p_2$, which leaves out a delta-function $\delta(p_2^2-m_2^2)$, where 
\begin{equation}
p_2^2 - m_2^2 = Q + 2(p_3\cdot p_4 - p_1\cdot p_3 - p_1\cdot p_4) \,,
\end{equation}
where $Q = m_1^2 - m_2^2 + m_3^2 + m_4^2$.  Measuring angles relative to the external momentum $p_1$, we define the coordinate system such that
\begin{align}
	\bm{p}_1 &= p_1 \left(0,0,1 \right) \,,
	\nonumber \\
	\bm{p}_3 &= p_3 \left(0,\sin\theta,\cos\theta \right)\,,
	\nonumber \\
	\bm{p}_4 &= p_4 \left(\sin\alpha\sin\beta,\sin\alpha\cos\beta,\cos\alpha \right)\,.
\end{align}
The remaining integrals can now be readily reduced to
\begin{align}
	{\mathrm{d}}^3\!p_3 &= 2\pi\, p_3^2{\mathrm{d}}p_3\,{\mathrm{d}}\cos\theta\, \,,
	\\
	{\mathrm{d}}^3\!p_4 &= p_4^2{\mathrm{d}}p_4\,{\mathrm{d}}\cos\alpha\,{\mathrm{d}}\beta \,.
	\nonumber
\end{align}
The integral over $\beta$ can be done~\cite{Hannestad:1995rs} using 
\begin{equation}
\delta(p_2^2-m_2^2) = \delta(f(\beta)) = \sum_i|f^\prime (\beta_i)|^{-1}\delta(\beta-\beta_i),
\end{equation}
where $f^\prime (\beta) = 2p_3p_4\sin\alpha\sin\theta\sin\beta$ and $\beta_i$ are the roots of $f(\beta)=0$. It is easy to see that the $\delta$ function is symmetric in $\beta$, so we can multiply by 2 and integrate over $[0,\pi ]$. Thus, $\cos\beta_i \in [-1,1]$ which demands $\sin^2\beta_i \geq 0$ for the root to exist within the integration interval. This can be equivalently written as
\begin{equation}
	f^\prime(\beta_i)^2
	= \left(2p_3p_4\sin\alpha\sin\theta\sin\beta_i\right)^2 \geq 0 \,,
\end{equation}
so that 
\begin{equation}
	\int_{0}^{2\pi} {\mathrm{d}}\beta  \, \delta \big( f(\beta) \big)
	= 2 |f^\prime (\beta_i)|^{-1}
	\Theta\!\left(|f^\prime (\beta_i)|^2 \right).
\end{equation}
Using the equation $f(\beta_i)=0$, one can rewrite the derivative as
\begin{equation}
	|f^\prime (\beta_i)|
	= \sqrt{a\cos^2\alpha + b\cos\alpha + c} \,,
\end{equation}
where, again paralleling the notation of \cite{Yueh:1976aa,Hannestad:1995rs},
\begin{align}
	a &= p_4^2(-4\kappa + 8\epsilon)\,,
	\label{eq:afact}
	\\
	b &= p_4(-p_1 + \epsilon/p_1)(8\gamma + 4Q + 8\epsilon) \,,
	\label{eq:bfact}
	\\
	c &= 4p_3^2p_4^2 \sin^2\theta - \big( 2(\gamma +\epsilon) + Q \big)^2 \,,
	\label{eq:cfact}
\end{align}
with $\gamma \equiv E_3E_4 - E_1E_3 - E_1E_4$ and $\epsilon \equiv p_1p_3\cos\theta$ and $\kappa \equiv p_1^2 + p_3^2$. Then, after a trivial ${\mathrm{d}}\mu$ integral, the source term becomes
\begin{align}
	\label{eq:BW_final}
	\mathcal{C}_\mathrm{BW}
	=
	\frac{2}{(2\pi)^4} \frac{1}{2E_1}
	&\iint \frac{ {\mathrm{d}}p_3 p_3^2 }{2E_3}
	\frac{ {\mathrm{d}}p_4 p_4^2}{2E_4}
	\,F(p_1,p_3,p_4) 
	\nonumber \\
	&\times f_3 f_4\left[1\pm f_1 \right]\left[1\pm f_2 \right]  \,,
\end{align}
where $f_i \equiv f_i(E_i)$ with $E_2=E_4+E_3-E_1$, and the function $F(p_1,p_3,p_4)$ contains the squared matrix element integrated over the angles and fixing kinematics,
\begin{align}
	F(p_1,p_3&,p_4) =
	\int_{-1}^{1} {\mathrm{d}}\cos\theta
	\int_{\max(-1,z^+)}^{\min(1,z^-)} {\mathrm{d}}z
	\nonumber \\
    &\times	
	\frac{|M(s,t)|^2} {\sqrt{a(z-z^-)(z - z^+)}}
	\Theta\!\left( b^2 - 4ac \right)\,,
	\label{eq:F_function}
\end{align}
where we set $z=\cos\alpha$ and $z^\pm \equiv (-b\pm\sqrt{b^2-4ac})/2a$. This function contains all process specific dynamical information and since it is independent of the distribution functions, it can be computed and fitted before solving the Boltzmann equations. In general, the matrix element squared is a function of $s = (p_3 + p_4)^2$ and $t=(p_1-p_3)^2$, which depend on the angles and momenta as follows:
\begin{align}
	t  =& m_1^2 + m_3^2 - 2E_1E_3 + 2p_1p_3\cos\theta \,,
	\nonumber \\
	s =& 2E_1E_3 + 2E_1E_4 - 2p_1p_3\cos\theta 
	\nonumber\\ 
	& - 2p_1p_4\cos\alpha + m_2^2 - m_1^2\,.
\end{align}
In the particular case where the matrix element in~\eqref{eq:F_function} does not depend on $\cos\alpha$ (a pure $t$-channel process), the $\mathrm{d}\cos$ integral can be reduced to a one-dimensional integral,
\begin{equation}
	\label{eq:F_pure_t}
	F(p_1,p_3,p_4)
	=
	\int\limits_{-1}^{1} {\mathrm{d}}\cos\theta
	\,|M(t)|^2 \frac{\pi}{\sqrt{-a}} \,\Theta\!\left( b^2-4ac\right)
	\,.
\end{equation}
%


\noindent{\it Forward term.}  In the forward term, given by Eq.~\eqref{eq:C_sink}, we integrate first over $\mathrm{d}^3p_4$, which leaves us with the delta function $\delta(p_4^2-m_4^2)$. Paralleling the backward term reduction, we eventually obtain
\begin{align}
	\label{eq:FW_final}
	\mathcal{C}_\mathrm{FW}
	= -\frac{2}{(2\pi)^4} \frac{1}{2E_1} &
	\iint \frac{ {p_2^2\mathrm{d}}p_2 }{2E_2}
	\frac{ {p_3^2\mathrm{d}}p_3 }{2E_3}\,
	F^{'}(p_1,p_2,p_3)
	\nonumber \\
	&\times f_1 f_2\left[1\pm f_3 \right]\left[1\pm f_4 \right],
\end{align}
where $f_4$ is evaluated at $E_4 = E_1+E_2-E_3$ and $F'(p_1,p_2,p_3)$ has an identical expression to the right-hand side of~\eqref{eq:F_function}, where one replaces everywhere $(a,b,c) \rightarrow (a',b',c')$ with
\begin{align}
	a' &= p_2^2(-4\kappa' + 8\epsilon')\,,
	\\
	b' &= p_2(p_1 - \epsilon'/p_1)(8\gamma' + 4Q' + 8\epsilon') \,,
	\\
	c' &= 4p_2^2p_3^2\sin^2\theta - \big( 2(\gamma' +\epsilon') + Q' \big)^2 \,,
\end{align}
where $Q'= m_1^2 + m_2^2 + m_3^2 - m_4^2$ and, moreover, $\gamma' = E_1E_2 - E_1E_3 - E_2E_3$ and $\epsilon' = p_1p_3\cos\theta$ and $\kappa' = p_1^2 + p_3^2$.

Now the Mandelstam variables must be written as $s=(p_1+p_2)^2$ and $t=(p_1-p_3)^2$ so that
\begin{align}
	s &= m_1^2 + m_2^2 + 2E_1E_2 - 2p_1p_2\cos\alpha,
	\nonumber\\
	t &= m_1^2 + m_3^2 - 2E_1 E_3 + 2p_1 p_3 \cos\theta.
\end{align}
If the matrix element is again independent of $\cos\alpha$, the result \eqref{eq:F_pure_t} applies also as such, after replacing $(a,b,c) \rightarrow (a',b',c')$. Note that the forward and backward collision integrals~\eqref{eq:BW_final} and~\eqref{eq:FW_final} are valid for general Bose-Einstein and Fermi-Dirac statistics.

\subsection{Special case: Inelastic 2-2 scattering toward/from equilibrium in MB statistics} 
\label{app:EQ}
%
As stated in the main text, we are using the MB statistics throughout. This is not needed for our computation of the elastic rates, but to keep the computation time associated with the large number of inelastic interactions with the SM states manageable. In this case, the final states are in equilibrium, and we can reduce the nine-dimensional integral down to one-dimension. Overall, using the MB statistics amounts to about 10\% error in the overall magnitude of the elastic collision integral~\cite{Dolgov:1992wf}, which should have but a very small effect on the final abundance. Indeed, we checked that scaling the elastic collision terms by a factor 0.9-1.1 caused only a $ 0.7$\% change in the final abundance. Then, working under the assumption $f \ll 1$ and enforcing the detailed balance, we can write the phase space factor \eqref{eq:phase_space_factor} as
\begin{equation}
	\Lambda(f_1,f_2) = f_1^{\textrm{eq}}f_2^{\textrm{eq}} - f_1 f_2 \,.
\end{equation}
The collision can then be written as
\begin{equation}
	\mathcal{C}_\I(p_1,t)
	= \frac{1}{2E_1} \! \int {\mathrm{d}}^3\!\tilde{p}_2
	\Lambda(f_1,f_2)
	\sum_n F_\I(s)\,\sigma_\I^{(n)}(s) \,,
\end{equation}
where $n$ runs over different equilibrium states, $\sigma_\I^{(n)}(s)$ is the corresponding cross section, and $F_\I(s)$ is the Lorentz invariant flux factor,
\begin{align}
	F_\I(s)
	&= 4\sqrt{(p_1 \cdot p_2)^2 -m_1^2m_2^2}\,,\nonumber \\
	&= 2 \lambda^{1/2}(s,m_1^2,m_2^2)
	\equiv 4E_1E_2 v_\Mol,
\end{align}
where $\lambda(a,b,c) = (a-b-c)^2 -4bc$ is the K{\"a}ll{\'e}n kinetic function. As explained in~\cite{Ala-Mattinen:2019mpa}, one can reduce the integral over ${\rm d}^3 \tilde p_2$ to a single integral over $s$,
\begin{equation}
	\mathcal{C}_I(p_1,t)
	= \frac{1}{2\pi^2}
	\int_0^\infty \! {\mathrm{d}}p_2 p_2^2 
	\Lambda(f_1,f_2) 
	\sum_n\left[ F_\I \sigma_\I^{(n)} \right],
\end{equation}
where
\begin{equation}
	\left[ \,\mathcal{O} \, \right](p_1,p_2)
	\equiv
	\frac{1}{16 p_1 p_2 E_1 E_2}
	\int_{s_-}^{s_+}
	{\mathrm{d}}s \, \mathcal{O}(s) \,.
\end{equation}
with $s_\pm = m_1^2 + m_2^2 + 2E_1E_2 \pm 2p_1p_2$.
Again, this function can be evaluated and fitted for each collision channel before one attempts to solve the dynamical Boltzmann equations, which gives a dramatic boost in numerical efficiency.

\subsection{1-2 decays and fusions} 

A similar reduction that was carried out above for the $2\leftrightarrow 2$ scatterings, can be performed for the  $1\leftrightarrow 2$ processes. We shall assume that either the decaying particle or the decay products are in thermal equilibrium.\\

\noindent {\it Decay from (fusion to) equilibrium} $f_A^\mathrm{eq} \leftrightarrow f_1 f_2$. Assume we are tracking the species labeled as $"1"$, while the $"2"$ species is arbitrary and $"A"$ species follows the equilibrium. Then,
\begin{equation}
\label{eq:eq-12}
	\mathcal{C}^{\rm eq}_{A-12} (p_1,t)
	=
	\frac{1}{2E_1}\!\int \!{\rm d}{\rm{PS}}_{A2}
	\, \Lambda\!\left(f_A^\textrm{eq},f_1,f_2\right) \, |M_{A-12}|^2,
\end{equation}
where the phase space integration is denoted by
\begin{equation}
  {\rm d}{\rm{PS}}_{A2}=\mathrm{d}^3\!\tilde{p}_A \mathrm{d}^3\!\tilde{p}_2
  (2\pi)^4\delta^{(4)}\!\big(p_A - p_1 - p_2\big)\,,
\end{equation}
and the distribution factor $\Lambda$ is given by
\begin{align}
	\Lambda
	& =
	f_A^{\textrm{eq}}\big[1 \pm f_1\big]\big[1 \pm f_2\big]
		- f_1 f_2\big[1 \pm f^{\textrm{eq}}_A\big] \,, \nonumber \\
	&\rightarrow
	f_1^\textrm{eq} f_2^\textrm{eq} - f_1 f_2 \,,
\end{align}
where we again assumed that $f \ll 1$ and applied the detailed balance for the equilibrium state. A similar procedure as in the previous section eventually gives
\begin{equation}
	\mathcal{C}^{\rm eq}_{\textrm{A}-12} (p_1,t)
    = \frac{|M_{\textrm{A}-12}|^2}{16\pi E_1 p_1}
	 \Big( f_1^\textrm{eq} \chi[f_2^\mathrm{eq}]- f_1 \chi[f_2]\Big) \,,
\end{equation}
where we used the fact that the matrix element for the decay process is a constant and defined
\begin{equation}
\chi[f] \equiv \int\limits_{0}^{\infty} {\rm d}p_2 \frac{p_2}{E_2}f(p_2)\Theta(1-|\cos\theta_0|) \,,
\end{equation}
with
\begin{equation}
\cos\theta_0 = \dfrac{|m_1^2 + m_2^2 - m_A^2 + 2E_1E_2|}{2p_1p_2}.
\end{equation}
In this case one only needs to compute the matrix element as a function of the masses of particles involved.\\

\noindent{\it Decay to (fusion from) equilibrium} 
$f_1 \leftrightarrow f_A^\mathrm{eq} f_{B}^\mathrm{eq}$. 
Now assume we track the species $"1"$ while the arbitrary species $"A"$ follows the equilibrium. After similar steps as above, we get an even simpler expression
\begin{align}
\label{eq:eq-12_v2}
	\mathcal{C}^{\rm eq}_{1\rightarrow AB} (p_1,t)
	=&
	\frac{|M_{1-AB}|^2}{16\pi E_1}\big(f_1^\textrm{eq} - f_1\big)
	v \Theta(v^2) \,,
\end{align}
where $v = \lambda^{1/2}(m_1^2,m_A^2,m_B^2)/m_1^2$ and
where we used the fact that the matrix element squared (in tree level) is always a constant.

%
\section{FEYNMAN RULES}
\label{app:feynmanrules}
%

Here we summarize the Feynman rules that follow from the Lagrangian in Eq.~\eqref{eq:symm_lagrangian} for the new/modified interaction vertices in the model with a singlet scalar $S=\sigma+w$ and a singlet fermion $N$. 
\begin{align}
\lambda_{\phi\phi\phi\phi} &= -6i\lambda_\H c_\theta^4 + 6s_\theta^2\left( \lambda_\HS c_\theta^2 + \lambda_\S s_\theta^2\right) ,
\nonumber \\
\lambda_{\sigma\sigma\sigma\sigma} &= 
-6i\lambda_\S c_\theta^4 + 6 s_\theta^2\left(\lambda_\HS c_\theta^2 + \lambda_\H s_\theta^2\right) ,
\nonumber \\
\lambda_{\phi\phi\sigma\sigma} &= 
-(i/4)\big(\lambda_\HS [1+3c_{4\theta}] + (3\lambda_\H + 3\lambda_\S) [1-c_{4\theta}]\big) ,
\nonumber \\
\lambda_{\phi\phi\phi} &= -6i\lambda_\H v c_\theta^3 
- 3 s_\theta \big(\lambda_\HS w c_\theta^2 - \lambda_\HS v s_\theta c_\theta + 2\lambda_\S w s_\theta^2\big) ,
\nonumber \\
\lambda_{\sigma\sigma\sigma} &= -6i\lambda_\S w c_\theta^3 
+ 3 s_\theta (\lambda_\HS v c_\theta^2 + \lambda_\HS w s_\theta c_\theta + 2\lambda_\H v s_\theta^2) ,
\nonumber \\
\lambda_{\phi\phi\sigma} &=-i\lambda_\HS v c_\theta^3 
- s_\theta (\lambda_{C2} w c_\theta^2 - \lambda_{C1} v s_\theta c_\theta + \lambda_\HS w s_\theta^2) ,
\nonumber \\
\lambda_{\phi\sigma\sigma} &=-i\lambda_\HS w c_\theta^3 + s_\theta (\lambda_{C1} v c_\theta^2 + \lambda_{C2} w s_\theta c_\theta + \lambda_\HS v s_\theta^2) ,
\nonumber \\
\lambda_{\phi VV} &= -2i(M_{V}^2/v) s_\theta g^{\mu\nu} \,,
\nonumber \\
\lambda_{\sigma VV} &= -2i(M_{V}^2/v) c_\theta g^{\mu\nu} \,,
\nonumber \\
\lambda_{\phi ff} &= -i(m_{f}/v) c_\theta \,,
\nonumber \\
\lambda_{\sigma ff} &= -i(m_{f}/v) s_\theta \,,
\nonumber \\
\lambda_{\phi NN} &= -i(y/2)s_\theta \,,
\nonumber \\
\lambda_{\sigma NN} & = \phantom{-}i(y/2)c_\theta \,.
\end{align}
Here $\phi$ denotes the SM Higgs field with VEV $v$ and $V$ and $f$ denote SM vector bosons and fermions; we used the following shorthand notations: $s_\theta\equiv \sin\theta$, $c_\theta\equiv\cos\theta$, where  $\theta $ is the mixing angle between the two scalars $\phi$ and $\sigma$, and finally $\lambda_{C1}\equiv 6\lambda_\H - 2\lambda_\HS $ and $\lambda_{C2}\equiv 6\lambda_\S - 2\lambda_\HS $.
%
\section{RIS SUBTRACTION} 
\label{app:RIS}
%

A well-known problem with Boltzmann equation networks concerns the double counting of pole contributions to scattering processes. The problem arises from imposing the on-shell limit for the dynamical two-point functions, when deriving the Boltzmann theory from the Kadanoff-Baym (KB) equations. In KB equations, where self-energies are defined by the two-particle irreducible (2PI)-effective action, the one-loop diagrams contain both on- and off-shell information. However, the on-shell limit spoils the 2PI hierarchy and removes the off-shell information at one-loop level. To regain it, one has to introduce by hand higher-order 1PI-self-energy diagrams, whose cuts then create most scattering processes in Boltzmann theory. This division does not  automatically respect unitarity and, in particular, the poles of the 2-2-scattering processes encode information already included in the fusion processes onto and in the decays of the unstable ``on-shell" state. Removing this double counting is usually dubbed real intermediate state (RIS) subtraction.

To see how this works, we split the Breit-Wigner propagator as follows:
\begin{align}
	iD(p^2) &\equiv 
	\frac{i}{p^2-m^2+im\Gamma} 
	\nonumber \\
	&= \frac{i(p^2-m^2)}{\big(p^2-m^2\big)^2+m^2\Gamma^2} 
	 + \frac{m\Gamma}{\big(p^2-m^2\big)^2+m^2\Gamma^2} 
	 	\nonumber \\
	&= iD_{\rm H}(p^2) + {\cal A}(p^2) \,,
\label{eq:Dp_full}
\end{align}
where $\Gamma$ is the decay width of the propagating particle with mass $m$. The propagator is resonant at $p^2=m^2$ with a resonance of width $m\Gamma$, isolated in the spectral function ${\cal A}(p^2)$. This observation clearly suggests to remove the on-shell contribution from the propagator by simply dropping the spectral part and retaining only the Hermitian part
\begin{equation}
	iD(p^2) 	\rightarrow iD_{\rm H}(p^2)
\label{eq:cline_presc}
\end{equation}
in all scattering calculations. This corresponds to removing the pole contribution at the level of the matrix element as proposed in~\cite{Cline:1993bd,Cline:2017qpe}. This is the approach taken in this paper. 

\begin{figure}[tb]
\hspace{-1em}
\includegraphics[width=0.42\textwidth ]{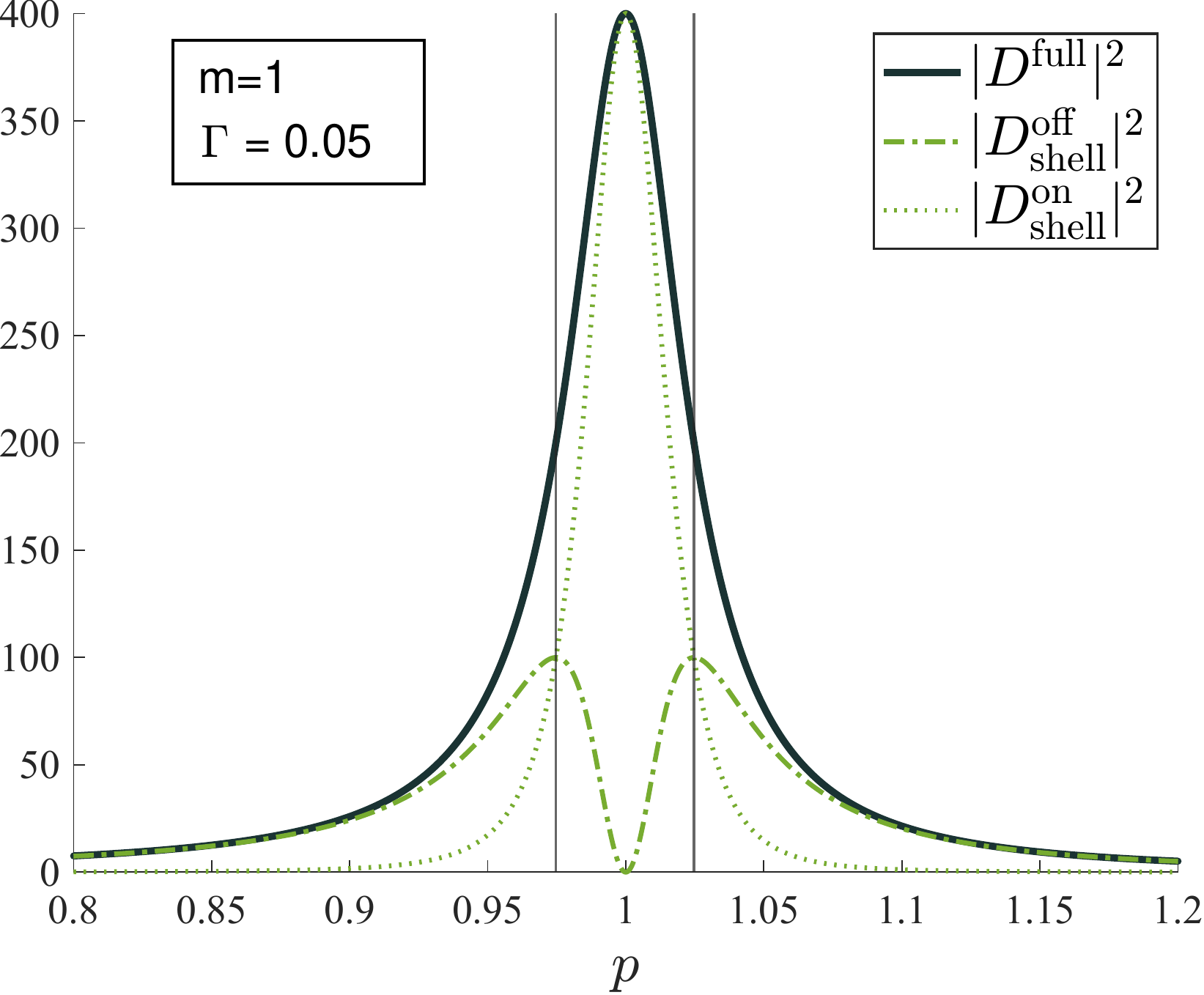}
\caption{ Shown are the Breit-Wigner propagator split into on- and off-shell part definitions used in \cite{Cline:2017qpe}. The split is done for the propagator and the resulting difference shown for the propagator squared. Full propagator $D(p^2)$ is shown with the black solid line, the off-shell part $D_{\rm H}(p^2)$ by the dash-dotted line, and the on-shell part ${\cal A}(p^2)$ by dash-dotted lines. The vertical lines mark the resonance width $p=\sqrt{m^2\pm m\Gamma}$, which coincides with the maxima of the off-shell propagator.}
\label{fig:CK_propagators}
\end{figure}

A common practice in the literature has been~\cite{Kolb:1979qa, Luty:1992un, Plumacher:1996kc,  Buchmuller:2000as,Giudice:2003jh,Pilaftsis:2003gt,Frigerio:2011in, Bernal:2018kcw} to subtract the divergence at the level of the matrix element squared, according to prescription
\begin{equation}
	|D(p^2)|^2 
	\rightarrow 
	|D(p^2)|^2  - \frac{\pi \delta(p^2-m^2)}{m\Gamma} \,.
\label{eq:stupid_definition}
\end{equation}
One can argue for this prescription by noting that, for a simple matrix element with no mixing between different channels,
\begin{equation}
|\mathcal{M}|^2 \sim D(p^2)D^*(p^2) = D_{\rm H}^2(p^2) + \mathcal{A}^2(p^2).
\end{equation}
To the lowest order in a small but finite $\Gamma$, the square of the spectral function can be replaced by
\begin{equation}
\mathcal{A}^2(p^2) \approx \frac{\pi \delta(p^2-m^2)}{m\Gamma}.
\end{equation}
The prescription~\eqref{eq:stupid_definition} is thus approximatively the same as~\eqref{eq:cline_presc}. However, this argument fails in the presence of interference terms and the prescription~\eqref{eq:stupid_definition} has been found to give superficial negative cross sections~\cite{Cline:1993bd}, which never happens in the subtraction scheme~\eqref{eq:cline_presc}. In Fig.~\ref{fig:CK_propagators} we show the split of the square of the propagator function into the on- and off-shell contributions according to~\eqref{eq:cline_presc} in a representative case.
\vspace{2em}
\FloatBarrier


\bibliography{main.bib}

\end{document}